\documentclass[11pt]{article}
\usepackage[margin=1.2in]{geometry}
\usepackage{graphicx}
\usepackage{booktabs}
\usepackage{threeparttable}
\usepackage{amsmath}
\usepackage{amssymb}
\usepackage{float}
\usepackage{natbib}
\usepackage[
  colorlinks=true,
  linkcolor=blue,
  citecolor=blue,
  urlcolor=blue
]{hyperref}
\usepackage{setspace}
\graphicspath{{figures/}}
\newcommand{\Description}[1]{}

\title{Public Trader Identity:\\ Adverse Selection and Return Predictability}
\author{Daojing Zhai\thanks{E-mail: \href{mailto:daojing.zhai@gmail.com}{%
\texttt{daojing.zhai@gmail.com}}. Homepage: \url{https://daojingzhai.github.io}.
GitHub repository: \url{https://github.com/daojingzhai/public-trader-identity}.}\\
Baruch College, City University of New York}
\date{August 2026}

\begin{document}
\onehalfspacing
\maketitle

\begin{abstract}
\noindent Informed traders are supposed to need anonymity: they profit by hiding among the
uninformed. A decentralized exchange now publishes the counterparty. Every committed order,
cancellation, rejection, and fill carries a persistent pseudonymous wallet address. We reconstruct
the full-depth limit order book from a record of 17.1~billion messages and 14.3~million aggressive
orders by 147,113 wallets, covering \$84.3~billion in taker notional. We report three findings. First, informativeness is a persistent wallet attribute. Wallets ranked by the price movement following their aggressive orders retain that ordering across adjacent ten-day windows, with a rank correlation of 0.52. Second, the ranking predicts returns. Adding the live activity of the highest-ranked wallets to a standard anonymous benchmark of prices, quotes, and order flow raises the out-of-sample $R^2$ for one-second returns to 12.31\%, a 13.2\% gain ($t=9.2$) that is 1.6 times the largest of 200 activity-matched placebo cohorts. Third, measured at realized trades rather than at every sampled moment, the increment grows from 1.43 to 2.47 percentage points of $R^2$. Public wallet histories therefore carry short-horizon price information that anonymous order-book data leave unmeasured.
\end{abstract} 


\section{Introduction}
\label{sec:introduction}

Anonymity is not an institutional detail in theories of informed trading; it is the condition under
which the mechanism operates. The informed trader earns by being indistinguishable from the flow
alongside which it trades, and because the maker cannot tell one counterparty from another, the
spread charges everyone for the losses caused by a few
\citep{bagehot1971only,kyle1985continuous,glosten1985bid}. Decentralized exchanges now intermediate
a growing share of derivatives volume, and on them that condition fails. Trading settles on a public
ledger under persistent addresses, so what an account has traded is visible to everyone, permanently
and at no cost. Two questions follow. First, do the same accounts stay expensive to trade against once everyone can
see who they are? Second, can tracking their activity sharpen a short-horizon price forecast beyond
what the anonymous book already gives? 

We answer these questions on Hyperliquid, the largest decentralized exchange for perpetual futures,
where the disclosure is unusually complete. It runs a
conventional central limit order book with price--time priority; what is unconventional is that the
book is maintained by a consensus protocol that publishes its entire input
\citep{albers2026openbook}. Every committed order submission, cancellation, rejection, and fill
becomes public when its block commits, carrying the persistent pseudonymous wallet address of the
account that sent it. We collect this record from a non-validating node and replay the messages in
consensus order, reconstructing the full-depth limit order book and linking each wallet's activity
through time. Across the ten most active perpetual markets, the July 2026 record contains
17.1~billion Level-4 messages and 14.3~million aggressive orders from 147,113 wallets, representing
\$84.3~billion in taker notional (Table~\ref{tab:data-summary}). The record answers both questions
directly, and lets us ask a third that only arises once they are settled: where in the trading
process the value of a wallet's history actually sits.

First, the same accounts do stay expensive to trade against. We score each wallet by the
notional-weighted ten-second markout of its aggressive orders over ten days and freeze the ranking
before evaluation; across adjacent ten-day windows the Spearman rank correlation is 0.52. The cost
is concentrated in a narrow tail: after adjusting for market, time, order size, volatility, and
spread, markouts remain flat through the fifteenth ventile and rise to 3.11 basis points in the
top ventile. We call the wallets at the top of this ranking toxic.

Second, tracking those accounts does sharpen the forecast. On a 100-millisecond grid, we compare a prespecified ten-variable benchmark of standard
predictors established in the literature, built from prices, quotes, and anonymous order flow,
with the same model
augmented by eleven features
constructed from previously scored toxic wallets. At the one-second headline horizon, identity
raises out-of-sample $R^2$ from 10.88\% to 12.31\%, a 13.2\% gain ($t=9.2$). The gain survives a broader wallet cohort, rolling and
peer-adjusted scores, longer information embargoes, and a nonlinear tree model.

Two exercises ask whether the ranking itself is doing the work. The first isolates it from size and
activity: we draw 200 cohorts of scored non-toxic wallets, matching each to the toxic cohort within
cells formed by deciles of scoring-window notional and order count, and evaluate all 201 cohorts on
the same grid, which yields 8.8~billion cohort-observations. At one second the toxic-wallet
increment is 1.6 times the largest matched draw and exceeds every draw through ten seconds. The
second applies the whole design, without retuning, to an independently collected public December
2025 sample covering 26.3~billion messages and 21.9~million taker fills by 89,800 wallets; the
prediction result recurs there (Appendix~\ref{app:dec}).

Third, the forecasts above are evaluated at every stamp on the 100-millisecond grid, and at most of
those stamps nothing trades at all. The natural question is whether identity still helps at the
fills that actually occur, which are the moments that cost a liquidity provider money. The increment is larger there: 2.47 percentage points of one-second
$R^2$ at an arrival, against 1.43 across the grid as a whole. That raises a second
question, because the simplest explanation for a larger gain at arrivals is that identity is a
readout of order flow: seeing historically toxic wallets resting at the quote indicates that a
trade is about to arrive and in which direction, so the result would reflect speed rather than
information in wallet histories. We examine that explanation in two ways. First, we reveal the
realized direction of the arriving order to both models, handing the forecaster exactly what a
readout of that order would supply. Second, we restrict the sample to fills after which no other
parent order trades, where there is no subsequent flow to read at all. We find that neither
removes the increment: revealing the direction absorbs 39\% of it, and what survives that also
survives the restriction. The signal is therefore not merely a reading of the arriving order's
direction, nor of the trades that follow it.

\paragraph{Related literature.}
This paper speaks first to the theory and measurement of adverse selection. Because the canonical
models leave the counterparty unidentified, the empirical literature infers adverse selection from
anonymous data: decompositions of the spread
\citep{glosten1988estimating,huang1997components}, the permanent price impact of trades
\citep{hasbrouck1991measuring}, the structurally estimated probability of informed trading
\citep{easley1996liquidity}, and its volume-clock successor \citep{easley2012flow}. These measures can fail exactly when identification
matters most:
measured adverse selection falls while genuinely informed investors trade
\citep{collindufresne2015do}. Persistent wallet tags add an observable account-level
dimension. We measure post-trade markouts by address and test their predictive content out of
sample.

Second, the paper contributes to the literature on trader identity and anonymity. Studies of venues that
concealed or revealed broker identifiers show that anonymity changes liquidity
\citep{foucault2007does,meling2021anonymous}, that order flow sorts across venues by anonymity,
though not always toward the anonymous one, since on the London Stock Exchange the informed
interdealer trades stayed in the direct, nonanonymous market \citep{reiss2005anonymity}, and that when identifiers are visible the market
learns from them \citep{linnainmaa2012lack}; in theory, credibly preannouncing an order can identify it as
uninformed and lower its execution cost \citep{admati1991sunshine}. Confidential account-level data show that performance
and information are persistent attributes of specific traders
\citep{barber2014crosssection,kirilenko2017flash,baron2019risk,vankervel2019high,hirschey2021do},
and knowing which client is behind a trade is valuable enough that brokers leaking it lets
their other clients trade against the liquidating fund
\citep{barbon2019brokers}. In our setting, a persistent identifier is attached to every
committed order and becomes public when its block is committed, rather than remaining visible
only to a broker, regulator, or researcher. We measure the identifier's value directly as an
out-of-sample forecasting increment.

Third, the paper adds to the literature on the microstructure of cryptocurrency and decentralized markets, which
has studied market quality and arbitrage \citep{makarov2020trading,barbon2025quality},
automated market makers, whose equilibrium liquidity provision has been compared directly with a
limit order book \citep{lehar2025decentralized} and whose passive liquidity loses to arbitrageurs
by construction \citep{milionis2022automated}, the strategic games public blockchains
enable \citep{daian2020flash}, and perpetual futures \citep{he2024fundamentals}. Public wallet
tags have also been used to study insider activity \citep{felezvinas2022insider}. Recent work
examines Hyperliquid's Level-4 record itself: \citet{albers2026openbook} introduce
the data, \citet{albers2026priceimpact} show that rejected orders predict returns, and
\citet{barone2026sunshine} show that disclosed execution intentions lower trading costs, an
on-venue test of \citet{admati1991sunshine}. A live maker experiment on the Binance perpetual
documents the fill-probability/adverse-selection trade-off \citep{albers2025dilemma}, and
market-by-order clustering recovers trader types from NASDAQ order events
\citep{zhang2025clusterlob}. We shift the unit of analysis to the persistent wallet: its past
markouts and its incremental forecasting content.

Finally, the results speak to market-making practice. Optimal quoting manages inventory and
adverse selection \citep{avellaneda2008high,cartea2015algorithmic}, conditions on short-horizon
book signals \citep{cartea2018enhancing}. Dealers can condition on client identity
\citep{cartea2025brokers,cartea2026detecting}, whereas anonymous forecasting models cannot
construct wallet histories from order-book data
\citep{cont2014price,sirignano2019universal,kolm2023deep}. Our forecasting comparison quantifies
the predictive content lost when public wallet histories are excluded.

\section{Institutional Setting and Data}
\label{sec:data}

Hyperliquid is the largest decentralized exchange for perpetual-futures trading. In July 2026,
it averaged \$6.6~billion in daily perpetual volume, accounting for 36.8\% of trading across
decentralized perpetual venues. The exchange runs on HyperCore, which maintains a central limit
order book for each contract and matches orders using price--time priority. Order submissions
and cancellations are sent to the network's validators, which collect them into short,
consensus-ordered \emph{blocks}; each block is then executed against the relevant order book.
Once a block is committed, its trades, positions, and resulting book state are final. The block
is therefore the atomic unit of ordering and information updates on Hyperliquid.

\begin{figure}[tbp]
    \centering
    \includegraphics[width=\textwidth]{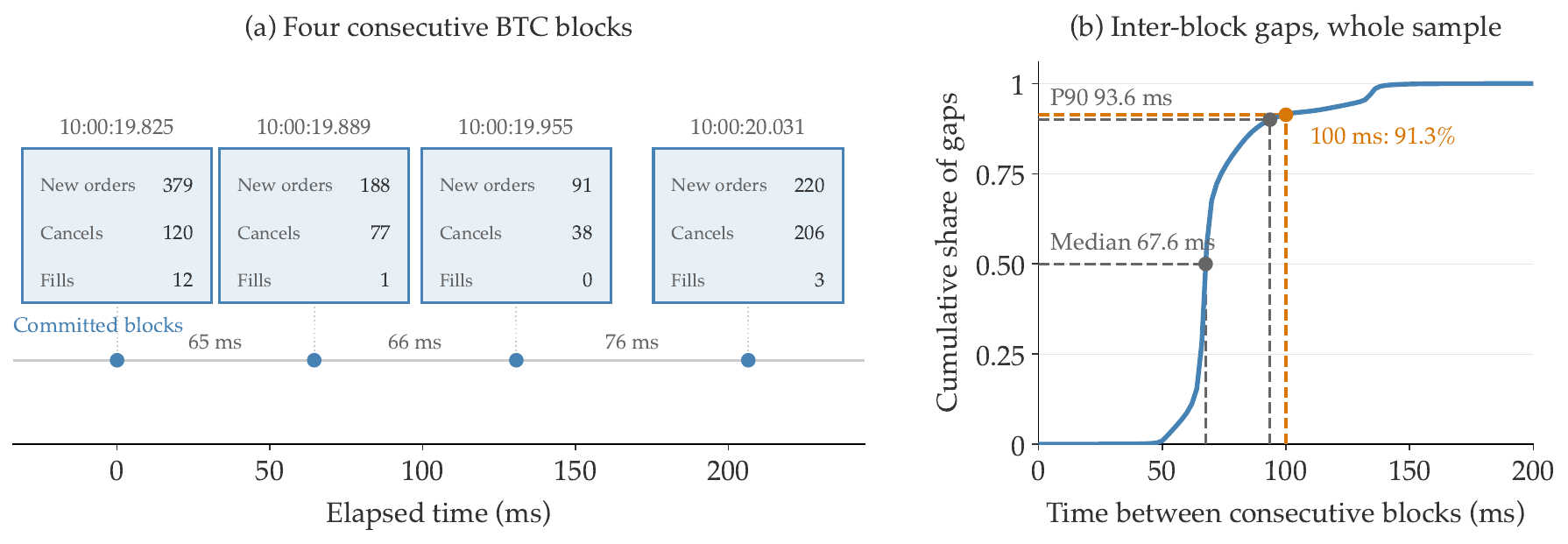}
    \caption{Committed blocks and inter-block timing. Panel (a) places four consecutive
    \textsc{btc} blocks on July 15, 2026, at their consensus timestamps; each block reports new
    orders, cancellations, and fills. Panel (b) shows the ECDF of all 22{,}734{,}866 gaps between
    adjacent blocks in the sample.}
    \label{fig:blocks}
\end{figure}

Because blocks are committed by consensus rather than at fixed wall-clock intervals, they arrive
on an irregular event clock. Figure~\ref{fig:blocks} makes this clock concrete. Four consecutive
\textsc{btc} blocks on July 15 illustrate what each block carries and how their spacing varies
(Panel~(a)). Across the full sample, the median time between adjacent blocks is 67.6 milliseconds,
the 90th percentile is 93.6 milliseconds, and 91.3\% of gaps do not exceed 100 milliseconds
(Panel~(b)). At any instant, the current midpoint is therefore the close of the most recent block
committed by that time. This committed-block clock underlies all prices and returns in the paper.

The distinctive feature of Hyperliquid is that all of this activity is public
\citep{albers2026openbook}. Every committed block records every order attempt (submissions,
cancellations, and rejections), together with every book update and execution. We collect these
records continuously through a non-validating node running on a dedicated cloud server. The node
connects to Hyperliquid's peer-to-peer network through bootstrap peers, synchronizes the committed
chain, and writes unbuffered output locally. In parallel, we query Hyperliquid's
\texttt{metaAndAssetCtxs} endpoint every ten seconds for mark prices, oracle prices, and open
interest; the oracle ticks provide an external check on the reconstruction. We then replay the
messages in consensus order to construct a Level-4 database of order-by-order book states,
reconstructed order lifecycles, rejected order attempts, matched maker and taker executions, and
the persistent pseudonymous wallet address associated with each order and trade.\footnote{We use
``identity'' throughout for this address-level behavioral identifier. One entity may control
several wallets, and one wallet may aggregate several principals; the analysis does not identify
legal persons.}
Appendix~\ref{app:reconstruction} provides details of the collection, reconstruction, and
validation procedures.

We apply this reconstruction to July 2026. Table~\ref{tab:data-summary} describes the ten
most active perpetual contracts by message count. The sample contains 17.1~billion Level-4
messages and 14.3~million aggressive orders from 147,113 wallets, representing 27.9~million
taker fills and \$84.3~billion in notional. Trading is led by \textsc{btc}, \textsc{eth}, and
\textsc{hype}, with \$46.8, \$18.7, and \$8.5~billion in taker notional, respectively. Our main
analyses focus on \textsc{btc}, \textsc{eth}, and \textsc{sol}, the three markets with the
greatest message activity. As a robustness check, Appendix~\ref{app:dec} replicates the main
analyses using the independently collected December 2025 data of
\citet{albers2026openbook}.

\begin{table}[!htbp]
\centering
\footnotesize
\begin{threeparttable}
\caption{Level-4 sample by perpetual market, July 1--27, 2026}
\label{tab:data-summary}
\setlength{\tabcolsep}{3.5pt}
\renewcommand{\arraystretch}{0.95}
\begin{tabular}{lrrrrrrr}
\toprule
Market & L4 messages & BBO updates & Agg. orders & Taker fills & Wallets & Notional & Spread \\
 & (bn) & (m) & (m) & (m) & (thousands) & (\$bn) & (bps) \\
\midrule
BTC & 9.75 & 0.94 & 4.02 & 8.66 & 98.9 & 46.8 & 0.20 \\
ETH & 3.03 & 0.91 & 2.04 & 3.89 & 61.7 & 18.7 & 0.62 \\
SOL & 1.35 & 1.69 & 1.35 & 2.16 & 50.5 & 4.9 & 0.26 \\
HYPE & 1.11 & 4.21 & 2.90 & 6.32 & 53.1 & 8.5 & 0.39 \\
ZEC & 0.66 & 6.78 & 1.16 & 2.06 & 26.3 & 2.4 & 0.75 \\
LIT & 0.33 & 4.08 & 1.11 & 1.95 & 14.1 & 0.9 & 2.62 \\
WLD & 0.32 & 4.73 & 0.43 & 0.71 & 18.7 & 0.4 & 2.14 \\
XRP & 0.28 & 0.79 & 0.32 & 0.54 & 21.3 & 0.7 & 1.01 \\
NEAR & 0.16 & 2.56 & 0.45 & 0.71 & 17.5 & 0.4 & 2.00 \\
PUMP & 0.09 & 0.58 & 0.51 & 0.89 & 20.9 & 0.5 & 7.13 \\
\midrule
All & 17.09 & 27.26 & 14.27 & 27.90 & 147.1 & 84.3 & 0.92 \\
\bottomrule
\end{tabular}
\begin{tablenotes}[flushleft]\footnotesize
\item \textit{Notes:} L4 messages combine raw order-status and book-diff records; BBO updates are
gate-valid blocks in which the best quote changed. Taker fills are individual taker legs. Because
one aggressive order can execute against several resting orders, fills that share a wallet,
order identifier, and market are grouped into one aggressive order. Spread is the mean quoted
bid--ask spread immediately before aggressive orders. Wallets are distinct within markets and
deduplicated across markets in the All row.
\item Markets are the ten most active Hyperliquid perpetuals by L4 message count; the All row
aggregates these markets.
\end{tablenotes}
\end{threeparttable}
\end{table}

\section{Wallet Toxicity}
\label{sec:toxicity}

Every action on Hyperliquid, including each order, cancellation, and fill, carries the pseudonymous
wallet address of the account behind it, allowing us to link a wallet's history to its subsequent
activity. Aggressive orders are particularly informative because the trader chooses immediate
execution and crosses the spread; canonical adverse-selection models therefore interpret
aggressive order flow as revealing beliefs about future prices
\citep{bagehot1971only,kyle1985continuous,glosten1985bid}. A directional trader may use aggressive
orders to express a view, while a market maker may use them to hedge inventory, respond to
order-flow imbalances, or act on short-term directional signals. Although these motives differ,
the direction of the resulting trade may contain information about subsequent prices. This
motivates our central hypothesis: wallets can be ranked by the signed markouts of their past
aggressive trades, and subsequent activity by wallets with historically high markouts should help
predict future returns.

This design requires persistence along two dimensions. First, wallet addresses must remain
observable long enough to accumulate a useful history. Of the 2{,}314 wallets with at least 100
qualifying aggressive orders during July 1--10, 86.9\% trade again during July 11--20. These
continuing wallets account for 50.3\% of all discretionary taker notional during July 1--10 and
43.6\% during July 11--20 across the \textsc{btc}, \textsc{eth}, and \textsc{sol} markets.
Second, informativeness itself must persist:
wallets whose aggressive trades are more informative in one period must remain more informative
in the next. In the remainder of this section, we first construct our measure of wallet
informativeness (Section~\ref{sec:measurement}) and then test whether it persists into a later
window (Section~\ref{sec:toxic-results}). Section~\ref{sec:prediction-results} then asks whether
the activity of informative wallets adds predictive content beyond the anonymous footprint of
current order flow.

\subsection{Measuring toxicity}
\label{sec:measurement}

Following the market-microstructure literature, we measure post-trade adverse selection using
the signed midpoint markout \citep{barone2026sunshine,cartea2026detecting}. For an aggressive
order $e$ filled at time $t_e$ with notional $n_e$ and direction $q_e$ ($+1$ buy, $-1$ sell),
the ten-second markout, in basis points of the pre-event midpoint, is
\begin{equation}
  x_e = 10^4\, q_e\,\frac{m_{t_e+10s}-m_{t_e^-}}{m_{t_e^-}},
  \label{eq:advselection}
\end{equation}
where $m_{t_e^-}$ is the midpoint at the close of the last block strictly before the event's
block, and $m_{t_e+10s}$ is the latest midpoint committed by $t_e+10s$. A positive markout means
that the midpoint moves in the aggressor's direction after the fill and therefore against the
passive counterparty. ``Toxicity'' throughout is a descriptive label for heterogeneity in this
quantity; it is not a claim that a wallet corresponds to one trader or trades on private
information. We use ten seconds as the benchmark horizon and report all markouts in
basis points; Figure~\ref{fig:persistence} also traces the markout from half a second to five
minutes.

Our benchmark wallet score is the notional-weighted mean markout over a frozen ten-day window,
\begin{equation}
  \alpha_w=\frac{\sum_{e\in w} n_e\, x_e}{\sum_{e\in w} n_e},
  \label{eq:wallet-toxicity}
\end{equation}
computed over July 1--10 for wallets with at least 100 qualifying aggressive orders; TWAP and
liquidation events are excluded. Freezing the score before model fitting and evaluation provides
a transparent out-of-sample benchmark. Among 2{,}314 scored wallets, the top decile (231
wallets) is our primary toxic cohort; the top quintile (462 wallets) is an alternative tail
cut.

The frozen window has one limitation: it gradually becomes stale and never scores wallets that
first become active after July 10. We therefore also re-estimate the score each day using the
previous ten days of trading (Appendix~\ref{app:score-alternatives}). A second alternative replaces
the raw markout with a peer-adjusted markout, defined as the order's markout net of other wallets'
notional-weighted mean markout in the same coin and minute (Appendix~\ref{app:peer-score}). This
adjustment removes price movements common to traders active at the same time. The persistence and
return-prediction results remain qualitatively unchanged across the frozen, rolling, raw, and
peer-adjusted definitions.

\subsection{Persistence, portrait, robustness}
\label{sec:toxic-results}

\begin{figure}[H]
    \centering
    \includegraphics[width=\textwidth]{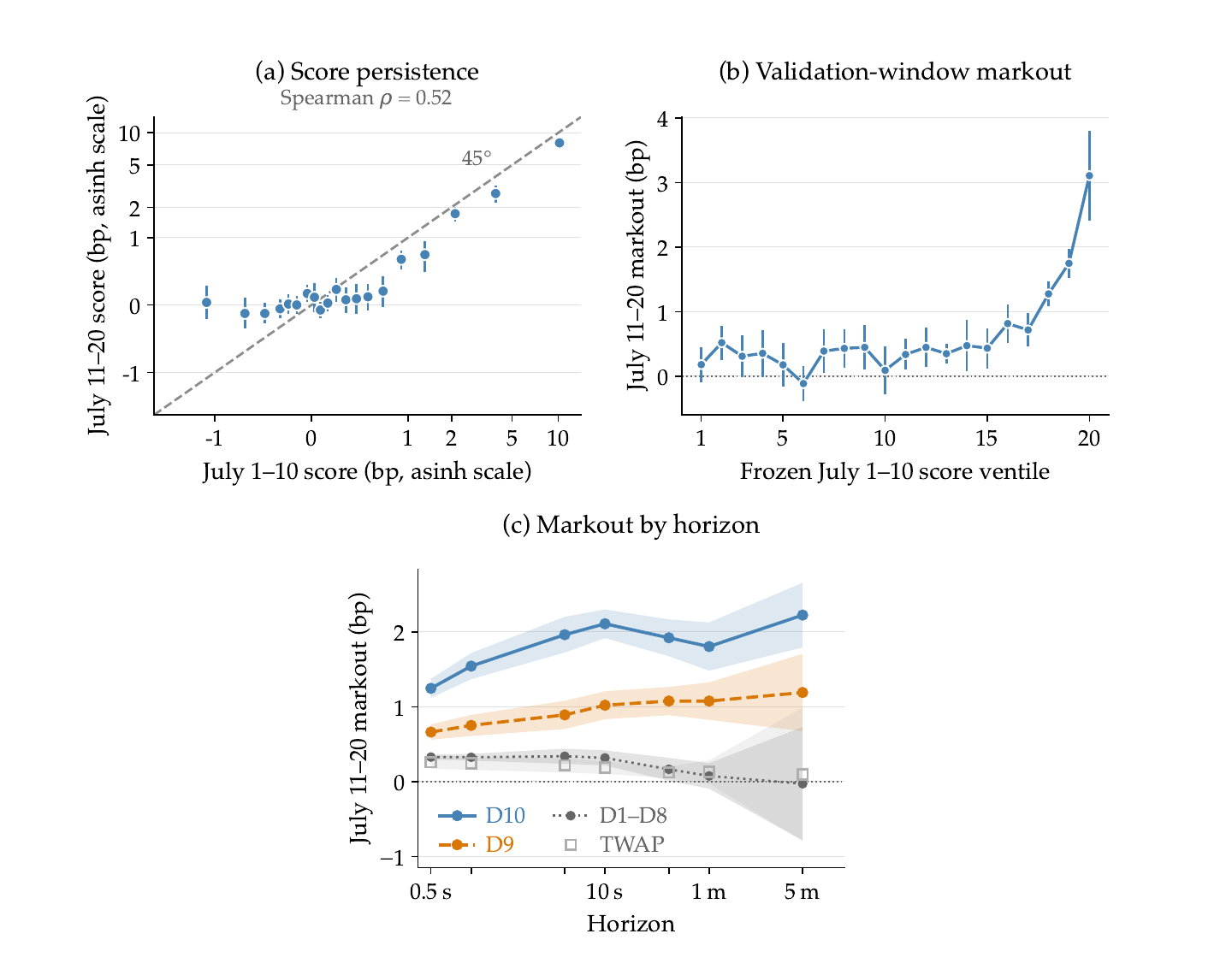}
    \caption{Persistence and horizon profile of wallet toxicity. Panel (a) compares scoring- and
    validation-window scores. Panel (b) reports adjusted validation markouts by scoring-window
    ventile. Panel (c) traces markouts for D10, D9, D1--D8, and native TWAP child orders. Bars and
    bands are 95\% confidence intervals.}
    
    \label{fig:persistence}
\end{figure}

Wallet informativeness is useful only if it persists out of sample. We use July 1--10 as the
scoring window and July 11--20 as the validation window. Our persistence hypothesis is that the
scoring-window ranking should continue to order wallets' validation-window markouts.
Figure~\ref{fig:persistence} supports this hypothesis. Among wallets that clear the 100-order
threshold in both windows, scoring- and validation-window scores remain strongly aligned (rank
correlation $\rho=0.52$). Panel~(a) shows that this persistence is asymmetric. The bottom decile
regresses toward zero, whereas the top decile remains elevated and close to the 45-degree line.
This upper-tail persistence is especially relevant because the prediction analysis tracks the
top decile. Because the score is a finite-sample average, $\rho=0.52$ understates the stability of
the underlying trait. Splitting each ten-day window into odd and even calendar days and rescoring
the same wallets on each half gives a split-half rank correlation of 0.52 for July 1--10 and 0.48
for July 11--20; the Spearman--Brown correction to full-window length puts the reliability ceiling
at 0.68 and 0.65, so the observed across-window figure attains roughly four-fifths of what a
perfectly stable trait could deliver and the implied ten-day decay in informativeness is modest.

The ventile sort provides a more direct test of whether high scoring-window ranks predict larger
validation-window markouts. After controlling for market, time, order size, volatility, and
spread, markouts remain flat through the fifteenth ventile before rising sharply to 3.11 basis
points in the top ventile (Panel~(b)). The economic content of wallet toxicity is therefore
concentrated in the upper tail.

A persistent ranking could still reflect temporary price pressure generated by aggressive
trades. If so, the markout should quickly reverse. Instead, the top-decile markout rises from
1.25 basis points at half a second to 2.11 basis points at ten seconds and remains elevated at
five minutes (Panel~(c)). The D1--D8 remainder stays near zero. Native TWAP child orders, whose
timing is scheduled rather than chosen separately for each execution, provide a benchmark for
mechanical execution. The D1--D8 remainder tracks this benchmark at every horizon, whereas the
top-decile markout is several times larger and rises as the TWAP markout decays. This horizon
profile is inconsistent with purely transient price pressure, although persistent impact,
repeated same-direction flow, and latent metaorders remain possible.

Persistence is economically relevant only if the top-ranked wallets continue to trade.
Table~\ref{tab:part1-panels} shows that 91.3\% of the 231 top-decile wallets trade again during
the validation window. Across the fixed deciles, validation-window markouts remain sharply ordered,
staying flat through D8, doubling in D9, and doubling again in D10. The result therefore reflects
persistence among continuing wallets rather than attrition among the initially ranked wallets.

\paragraph{Cross-venue effects.}
Persistent markouts could arise mechanically if some wallets react to price changes on other
venues before Hyperliquid adjusts, the cross-venue analogue of toxic arbitrage
\citep{foucault2017toxic}. Such a wallet should trade toward the gap between the venue's
external price oracle and its local midpoint, buying when the oracle is higher and selling when
it is lower. Table~\ref{tab:latency} in Appendix~\ref{app:latency} shows the opposite pattern.
Top-decile orders
arrive on the gap's side only 41.3\% of the time, compared with 57.9\% for the bottom decile, and
the mean signed gap they trade into is negative. The markout also survives two controls for this
gap. Adding gap controls moves the daily D10--D1 coefficient only from 1.72 to 1.75, while the
subsample with a gap no larger than one basis point still yields 2.12 basis points for the top
decile and 0.16 for the bottom decile. Together, these results show that cross-venue price gaps
do not explain the top-decile markout.

\begin{table}[H]\centering
\begin{threeparttable}
\caption{Wallet deciles: scoring window and out-of-sample validation}
\label{tab:part1-panels}
\footnotesize\setlength{\tabcolsep}{2.9pt}
\begin{tabular}{lrrrrrrrrrr}
\toprule
 & \multicolumn{5}{c}{Scoring window (July 1--10)} & \multicolumn{5}{c}{Validation window (July 11--20)} \\
\cmidrule(lr){2-6} \cmidrule(lr){7-11}
 & Wallets & Orders & Markout & Median & Notional & Wallets & Orders & Markout & Median & Notional \\
Decile & & & (bps) & order (\$) & share (\%) & & & (bps) & order (\$) & share (\%) \\
\midrule
D1 & 231 & 53,758 & $-$1.13 & 83 & 0.5 & 203 & 36,380 & 0.27 & 59 & 0.4 \\
D2 & 232 & 74,444 & $-$0.39 & 47 & 0.6 & 195 & 57,109 & 0.06 & 24 & 0.4 \\
D3 & 231 & 73,756 & $-$0.20 & 150 & 0.8 & 203 & 79,110 & $-$0.10 & 76 & 0.8 \\
D4 & 232 & 96,900 & $-$0.07 & 47 & 0.7 & 201 & 61,826 & 0.29 & 46 & 0.4 \\
D5 & 231 & 89,907 & 0.06 & 74 & 0.6 & 193 & 45,794 & 0.03 & 81 & 0.3 \\
D6 & 232 & 113,316 & 0.17 & 200 & 2.2 & 199 & 86,027 & 0.24 & 361 & 1.7 \\
D7 & 231 & 72,503 & 0.32 & 982 & 2.5 & 196 & 40,285 & 0.30 & 790 & 1.7 \\
D8 & 232 & 77,536 & 0.54 & 428 & 2.3 & 205 & 36,803 & 0.60 & 357 & 1.7 \\
D9 & 231 & 263,740 & 0.96 & 276 & 11.9 & 204 & 180,469 & 1.09 & 500 & 11.0 \\
D10 & 231 & 596,480 & 2.56 & 1,098 & 31.0 & 211 & 408,061 & 2.20 & 1,201 & 25.1 \\
\bottomrule
\end{tabular}
\begin{tablenotes}[flushleft]\footnotesize
\item \textit{Notes:} Deciles are formed once, on the frozen July 1--10 wallet score, and
never re-sorted. Both windows report the identical five statistics of each decile's aggressive
orders on the three model markets (TWAP and liquidation events excluded): distinct wallets
trading, aggressive-order count, notional-weighted ten-second markout, median order notional,
and the decile's share of all discretionary taker notional in that window. The share column is
therefore taken against the whole window's taker flow rather than the scored subtotal, so it does
not sum to one hundred percent; the residual is flow from addresses without a qualifying
scoring-window history.
\end{tablenotes}
\end{threeparttable}\end{table}

\section{High-frequency Return Prediction with Identity}
\label{sec:prediction}

Section~\ref{sec:toxic-results} shows that post-trade markouts persist across identifiable wallets.
This section asks whether identifying the wallets behind market activity improves short-horizon
return forecasts beyond the information available from an anonymous order book. Our design
compares an anonymous benchmark with an identity-augmented model. Both observe the same public
tape. The benchmark summarizes aggregate prices, quotes, and trade flow, while the augmented
model adds parallel quote and flow measures for top-decile wallets, together with wallet-specific measures of trading breadth and quoted depth. The difference in predictive performance measures the incremental information revealed by wallet identity.

To make the comparison demanding, the anonymous block combines well-established predictors from
several strands of the market-microstructure literature. Table~\ref{tab:part2-variables} maps each
variable to its source and groups the predictors into three families. Quote variables capture
depth imbalance at the best quotes and within five basis points of the midpoint, together with
quote-update order-flow imbalance over the previous one and thirty seconds
\citep{stoikov2018micro,xu2019multilevel,cont2014price}. Because the price move associated with a
given near-touch imbalance scales with volatility \citep{kyle1985continuous}, we multiply the
near-touch pair by the coin's realized one-second return volatility on the previous day and
express both variables in basis points of implied price movement. Flow variables capture the
latest signed print and signed taker notional over the same windows
\citep{hasbrouck1991measuring,biais1995empirical}. Price variables capture the previous one-minute
return, one-minute realized volatility, and quoted spread
\citep{andersen2003modeling,glosten1988estimating}.

The identity block applies the same construction to quote and flow activity from top-decile
wallets. It also includes variables that can be measured only when wallets are distinguishable:
the number of distinct top-decile buyers minus sellers, the decile's share of best-quote depth,
and its contribution to best-quote imbalance. Price variables have no identity counterpart
because prices are common to all traders.\footnote{The matched-wallet placebo asks whether knowing
the identities of an equally active but non-toxic wallet cohort would produce the same gain. We
repeat the identity construction for 200 cohorts matched to the toxic wallets on scoring-window
order count and notional; none reproduces the toxic cohort's gain through ten seconds
(Appendix~\ref{app:matched-placebo}).}

\begin{table}[tbp]\centering
\begin{threeparttable}
\caption{Predictor sets: the public--identity mapping}
\label{tab:part2-variables}
\footnotesize\setlength{\tabcolsep}{4.5pt}
\begin{tabular}{lll}
\toprule
Family & Variable (code) & Reference \\
\midrule
\multicolumn{3}{l}{\textit{Panel A: paired variables (identical construction; identity restricts to the toxic decile)}} \\
\midrule
Quote & Best-quote depth imbalance (\texttt{imb} / \texttt{hotImb}) & \cite{stoikov2018micro} \\
Quote & Near-touch depth imbalance\tnote{c} (\texttt{nearImb} / \texttt{hotNear}) & \cite{xu2019multilevel} \\
Quote & Quote-update order-flow imbalance, 1s (\texttt{ofi1s} / \texttt{hotOfi1s}) & \cite{cont2014price} \\
Quote & Quote-update order-flow imbalance, 30s (\texttt{ofi30s} / \texttt{hotOfi30s}) & \cite{cont2014price} \\
Flow & Latest signed print (\texttt{print} / \texttt{hotPrint}) & \cite{hasbrouck1991measuring} \\
Flow & Signed taker notional, 1s (\texttt{flow1s} / \texttt{hotFlow1s}) & \cite{biais1995empirical} \\
Flow & Signed taker notional, 30s (\texttt{flow30s} / \texttt{hotFlow30s}) & \cite{biais1995empirical} \\
\midrule
\multicolumn{3}{l}{\textit{Panel B: public only (prices are common to all observers)}} \\
\midrule
Price & One-minute midpoint return (\texttt{ret1m}) & \cite{andersen2003modeling} \\
Price & One-minute realized volatility of 1s returns (\texttt{vol1m}) & \cite{andersen2003modeling} \\
Price & Quoted spread, bp of mid (\texttt{spread}) & \cite{glosten1988estimating} \\
\midrule
\multicolumn{3}{l}{\textit{Panel C: identity only (no counterpart in the anonymous block)}} \\
\midrule
Flow & Net distinct toxic-decile buyers, 1s\tnote{a} (\texttt{hotBreadth1s}) &  \\
Flow & Net distinct toxic-decile buyers, 30s\tnote{a} (\texttt{hotBreadth30s}) &  \\
Quote & Toxic-decile share of best-quote depth\tnote{b} (\texttt{hotShare}) &  \\
Quote & Toxic-decile contribution to best-quote imbalance\tnote{b} (\texttt{hotContrib}) &  \\
\bottomrule
\end{tabular}
\begin{tablenotes}[flushleft]\footnotesize
\item \textit{Notes:} Parentheses give the code identifiers used in the estimation and in
Figure~\ref{fig:dropone}: in Panel A the anonymous variable first and its identity twin
second, separated by a slash. Panel A pairs share a window, formula, and clock, the
\texttt{hot} prefix restricting the construction to the toxic decile; Panel C variables
exist only because wallets are distinguishable, so they carry no citation.
\item[a] Counts of distinct wallets; the anonymous block measures flow by signed notional rather than by trader count.
\item[b] Scaled by total best-quote depth: the decile's weight in the book, not its internal tilt.
\item[c] Depth within 5bp of the mid, in basis points of implied move (Section~\ref{sec:prediction}).
\end{tablenotes}
\end{threeparttable}\end{table}

\subsection{Design and Freezing Protocol}
\label{sec:prediction-method}

Return forecasting requires fixed calendar horizons, whereas Hyperliquid updates on an irregular
block clock. A block-to-block target would therefore mix returns measured over different elapsed
times. We instead sample the market on a regular 100-millisecond grid. This interval remains close
to the venue's update frequency because 91.3\% of adjacent blocks are committed within 100
milliseconds (Figure~\ref{fig:blocks}, Panel~(b)).

Let $t$ denote a grid timestamp. Every feature uses only information committed by $t$. Book and
price variables are states, so they use the close of the latest block committed before
$t$. Trade flow and quote-update order-flow imbalance are interval measures, so their values at
$t$ aggregate events in the completed interval $[t-100\text{ ms},t)$. Recording the interval only
at its endpoint prevents the data leakage that would arise were events after the forecast time to enter the
feature vector.

For coin $c$ and horizon $\Delta$, the target is the forward midpoint return
\[
y_{c,t}(\Delta)=\frac{m_{c,t+\Delta}-m_{c,t}}{m_{c,t}},
\]
where each midpoint is the close of the latest block committed by the corresponding grid
timestamp.\footnote{The same block-close rule governs Equation~\eqref{eq:advselection}. Its
pre-event midpoint uses the last block strictly before the event's block and must be at most 60
seconds old. The endpoint uses the last block committed by $t_e+h$ and is missing if that state is
more than 120 seconds old, outside the sample, or inside a gate-suppressed interval. The prediction
grid carries the latest emitted state forward and does not apply these markout-specific staleness
screens. All timing is indexed by consensus timestamps on the committed tape rather than by a
separately audited local receipt-time clock.} We report horizons from 200 milliseconds to thirty
seconds and set targets that cross UTC-day boundaries to missing. As a robustness check to further avoid data leakage, we shift
the entire feature vector back by an additional 200 and 300 milliseconds while leaving the target
at $t$.

The protocol freezes everything in sequence: wallet scores on July 1--10; model fitting on
July 11--20 (ridge with coin intercepts as the benchmark; features
standardized with fitting-window moments; penalties by expanding-day validation); one
evaluation pass on July 21--27.\footnote{The window ends July 27 because the collection
infrastructure, not the market, sets the boundary: the collector fell behind the chain during
July 27 and the node failed outright at 06:06:45 UTC on July 28. This documented infrastructure
event was fixed independently of any model output and is detailed in
Appendix~\ref{app:collection}.} We report
$R^2=1-\sum_{c,t}(y_{c,t}-\widehat y_{c,t})^2/\sum_{c,t}y_{c,t}^2$ against a zero forecast,
equal-weighted per grid cell, with inference from the seven paired daily out-of-sample MSE
differences.

\subsection{The Incremental Value of Identity}
\label{sec:prediction-results}

If wallet histories contain information absent from the anonymous tape, adding identity should
improve forecasts on dates untouched by model selection. Consistent with this hypothesis, the
identity-only model produces an out-of-sample $R^2$ of 10.06\% at one second under the frozen
score, compared with 10.88\% for the anonymous block, even though the identity features observe
only the toxic decile's activity.
Combining the blocks raises $R^2$ to 12.31\%, a 13.2\% relative gain ($t=9.2$).
Table~\ref{tab:part2-v1} shows that this improvement is not confined to the headline horizon.
The gain remains distinguishable from zero through thirty seconds. We limit the reported horizon
to thirty seconds because the anonymous $R^2$ subsequently falls toward 2\%, making relative
gains on the small denominator difficult to compare with the short-horizon results.

Gradient-boosted trees extract substantially more information from the anonymous features,
raising their one-second $R^2$ to 19.48\%. Even against this stronger nonlinear benchmark,
identity raises $R^2$ to 20.65\%, a 6.0\% relative gain ($t=5.0$). The smaller gain than under
ridge indicates that nonlinearities in the anonymous features capture part, but not all, of the
information supplied by identity. The gain under trees declines from 10.7\% at 200 milliseconds
to 2.7\% at thirty seconds. The identity gain remains distinguishable from zero through ten
seconds, but not at thirty seconds ($t=1.6$).

The result is also insensitive to how wallet scores and the toxic cohort are defined.
Re-estimating scores on a rolling ten-day window produces a 12.7\% one-second gain, close to the
13.2\% gain under the frozen score (Table~\ref{tab:part2-v1-appx}). Expanding the cohort from the
top decile to the top quintile leaves the one-second gain essentially unchanged, indicating that
the predictive content is concentrated in the top decile.

\begin{table*}[t]\centering
\begin{threeparttable}
\caption{Return prediction, top-decile identity block}
\label{tab:part2-v1}
\footnotesize\setlength{\tabcolsep}{5pt}
\begin{tabular}{lrrrrrrrr}
\toprule
 & \multicolumn{4}{c}{Ridge (benchmark)} & \multicolumn{4}{c}{Gradient-boosted trees} \\
\cmidrule(lr){2-5} \cmidrule(lr){6-9}
Horizon & Anonymous & + Identity & Gain (\%) & $t$ & Anonymous & + Identity & Gain (\%) & $t$ \\
 & \multicolumn{2}{c}{$R^2$ (\%)} & & & \multicolumn{2}{c}{$R^2$ (\%)} & & \\
\midrule
0.2s & 6.91 & 8.20 & +18.7 & 8.1 & 19.26 & 21.32 & +10.7 & 5.4 \\
0.5s & 10.16 & 11.76 & +15.8 & 8.3 & 21.00 & 22.94 & +9.2 & 4.6 \\
1s & 10.88 & 12.31 & +13.2 & 9.2 & 19.48 & 20.65 & +6.0 & 5.0 \\
2s & 10.60 & 11.66 & +10.0 & 8.5 & 16.35 & 17.04 & +4.2 & 6.0 \\
5s & 9.17 & 9.79 & +6.7 & 9.5 & 12.00 & 12.36 & +3.0 & 6.1 \\
10s & 7.18 & 7.56 & +5.2 & 9.9 & 8.55 & 8.76 & +2.5 & 6.4 \\
30s & 3.56 & 3.67 & +3.0 & 6.1 & 3.74 & 3.84 & +2.7 & 1.6 \\
\bottomrule
\end{tabular}
\begin{tablenotes}[flushleft]\footnotesize
\item \textit{Notes:} Frozen wallet score, top decile---the paper's headline specification.
Forward midpoint returns on the 100ms grid, three model markets; scores July 1--10, models fit
July 11--20, evaluation July 21--27, with trees trained July 11--19 and July 20 held out for
early stopping.
\item \emph{Anonymous} is the benchmark without identity features and \emph{+ Identity}
adds the identity block; both learners use the same variable blocks. The \emph{Anonymous} and
\emph{+ Identity} columns report out-of-sample $R^2$, Gain
$=R^2_{\mathrm{+Identity}}/R^2_{\mathrm{Anonymous}}-1$ is the relative increase, and $t$
comes from the seven paired daily MSE differences. Horizons past thirty seconds are omitted, the
anonymous $R^2$ falling toward two per cent there.
\end{tablenotes}
\end{threeparttable}\end{table*}

\begin{figure}[H]
    \centering
    \includegraphics[width=\textwidth]{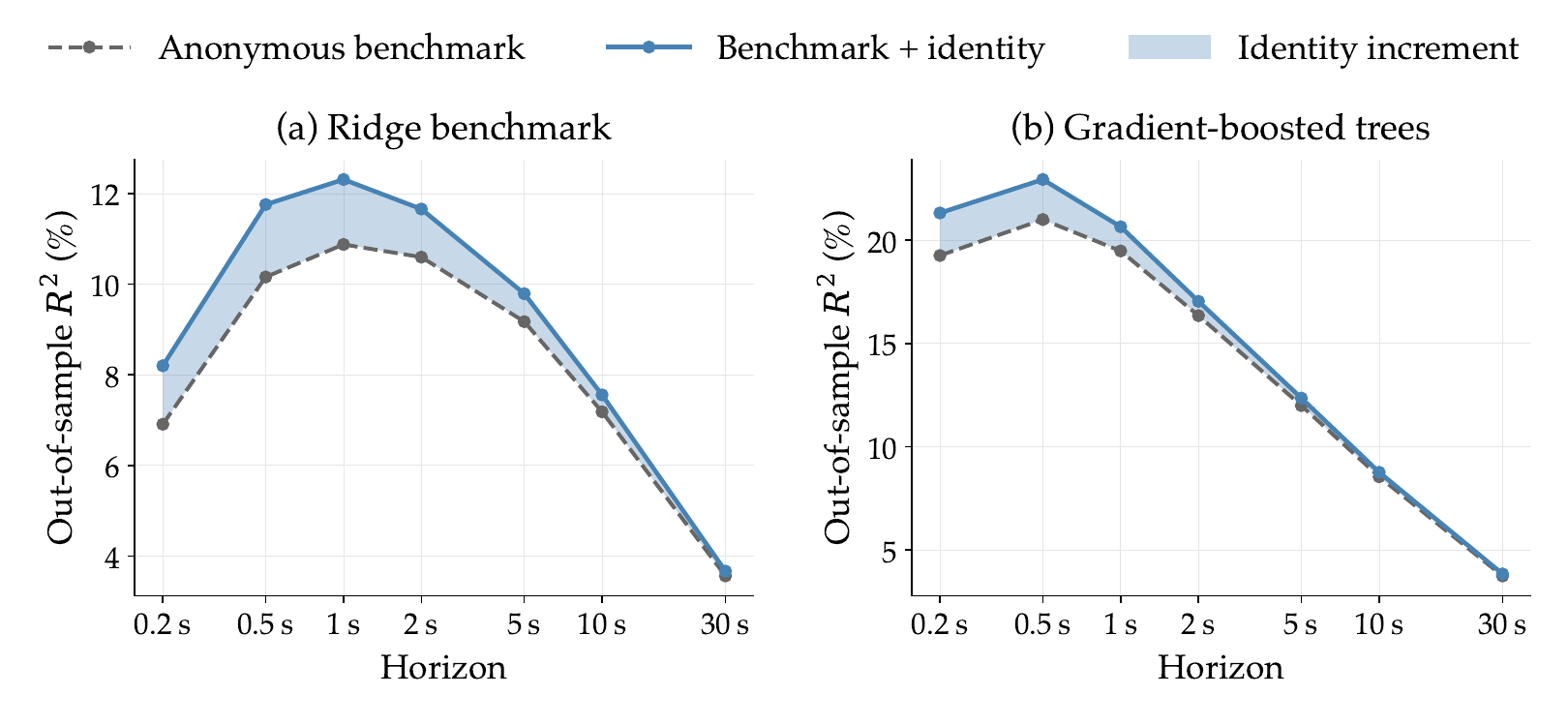}
    \caption{Out-of-sample $R^2$ by horizon under ridge and gradient-boosted trees. Gray is the anonymous benchmark, blue adds the top-decile identity block, and shading is the identity increment.}
    \label{fig:gain-decay}
\end{figure}

Order-flow toxicity should matter most immediately after information enters the market and fade
as the market absorbs it. The horizon profile in Figure~\ref{fig:gain-decay} supports this view.
Overall forecast fit is mildly hump-shaped, rising over the first few hundred milliseconds,
peaking between 0.5 and one second, and declining thereafter. This short effective horizon is
consistent with \citet{kolm2023deep}, who find that stock-specific order-book forecasts remain
effective for approximately two average price changes. The incremental value of identity decays
more steadily: its relative gain falls from 18.7\% to 3.0\% under ridge and from 10.7\% to 2.7\%
under trees between 200 milliseconds and thirty seconds. Identity therefore captures
fast-moving information rather than a persistent wallet characteristic that happens to correlate
with returns.

\paragraph{Where the identity value lives.}

\begin{figure}[H]
    \centering
    \includegraphics[width=\textwidth]{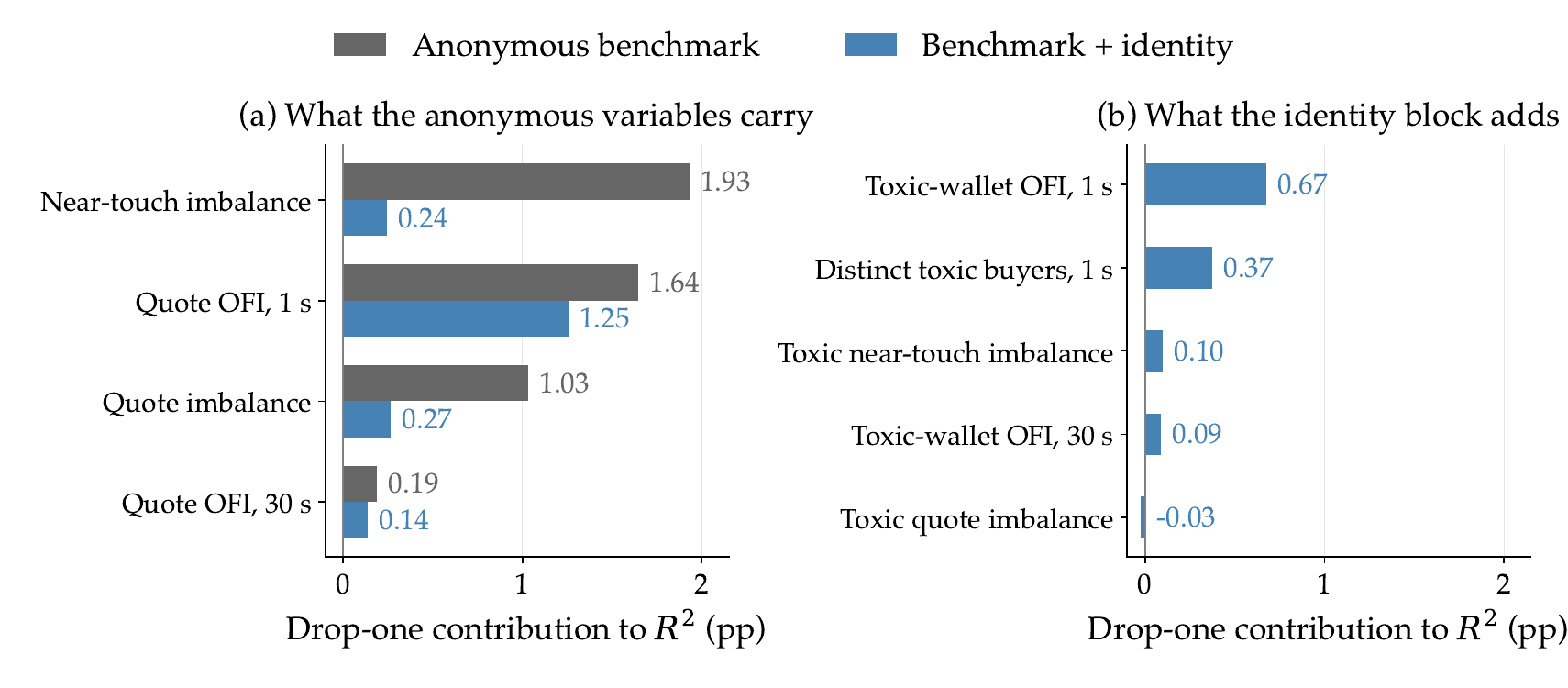}
    \caption{Drop-one contributions at the one-second ridge benchmark. Panel (a) compares anonymous predictors before and after adding identity; Panel (b) shows identity predictors. Contributions are descriptive and non-additive; each anonymous variable is shown with its identity counterpart, and pairs in which neither member reaches an $R^2$ contribution of 0.05 percentage points are omitted.}
    \label{fig:dropone}
\end{figure}

The aggregate gain could arise from a new identity-specific signal or from identity variables
substituting for anonymous predictors. Figure~\ref{fig:dropone} distinguishes these possibilities
by removing each variable in turn, refitting the one-second ridge model, and recording the loss
in out-of-sample $R^2$, in percentage points. Because the regressors are correlated, these
contributions are
descriptive and non-additive. In the anonymous benchmark, the largest contribution comes from
near-touch depth imbalance (\texttt{nearImb}, 1.93); once identity is added its
contribution falls to 0.24. The identity-side contributions are concentrated in the toxic
decile's one-second quote flow (\texttt{hotOfi1s}, 0.67) and net buyer breadth
(\texttt{hotBreadth1s}, 0.37). This pattern is consistent with overlap between anonymous
near-touch imbalance and the cohort-specific quote variables, not a causal decomposition.

\paragraph{Matched-wallet placebo.}

Toxic wallets are unusually large and active, so identity features built from any comparably
active cohort might improve forecasts. To isolate the toxicity ranking from size and activity,
we construct 200 cohorts from scored non-toxic wallets, matching each draw to the toxic cohort on
deciles of scoring-window notional crossed with order count. We then rebuild all eleven identity
variables for every cohort from the block-level book and flow records and carry them to the same
100-millisecond grid. In total, the exercise evaluates the real and 200 matched cohorts at
44.0~million grid stamps across fifty-one coin-days, yielding 8.8~billion cohort-observations
from 126~gigabytes of replayed per-wallet quote tapes.

At the one-second headline horizon, the toxic-decile identity increment is 1.6 times the maximum
among the 200 matched cohorts. The toxic cohort exceeds every draw at every horizon through ten
seconds (rank $p=1/201$). Figure~\ref{fig:placebo-hist} shows this separation for both tail cuts,
and Table~\ref{tab:p2placebo} in Appendix~\ref{app:matched-placebo} reports the exact increments
and ranks. At thirty seconds, the top decile enters the placebo range, while the top quintile
continues to exceed every draw. The toxicity ranking therefore adds information beyond size and
activity precisely at the horizons where identity is predictive.

\begin{figure}[H]
    \centering
    \includegraphics[width=\textwidth]{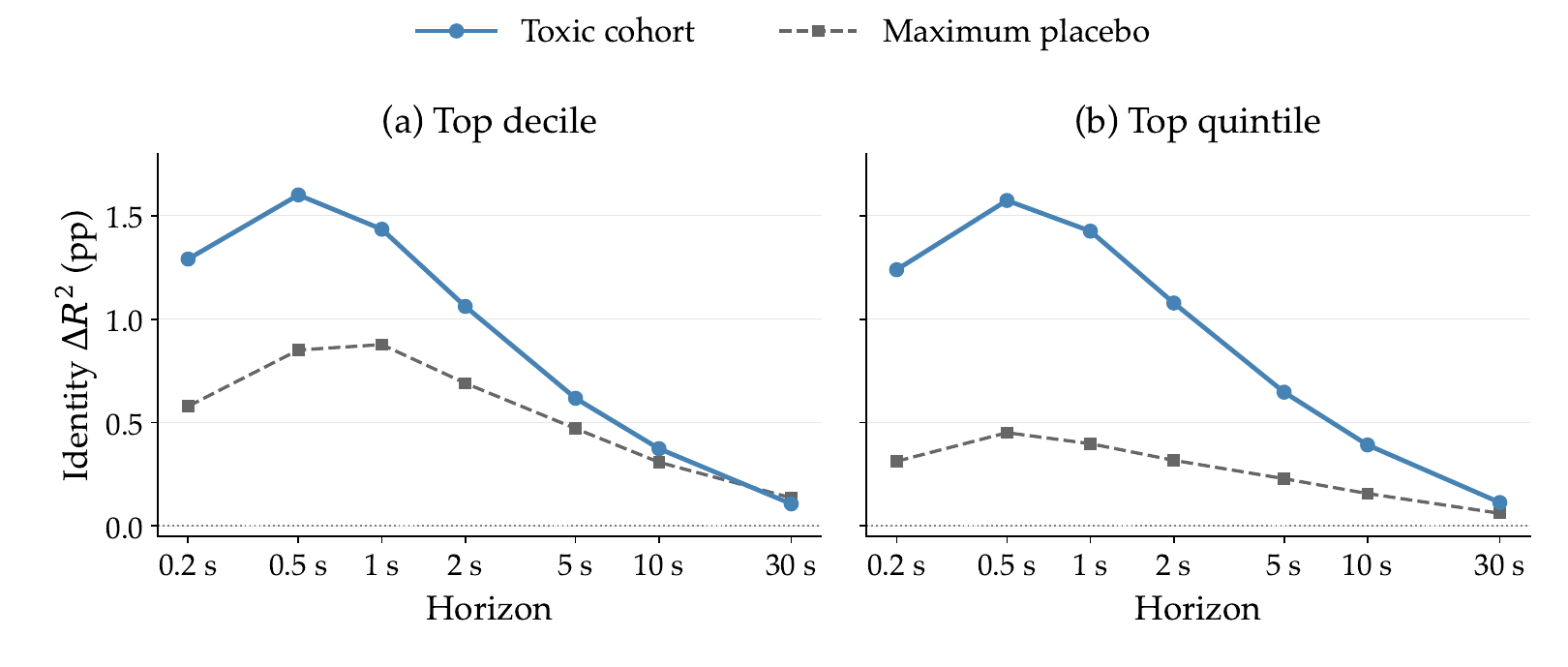}
    \caption{Identity increment for the toxic cohort against the maximum across 200 activity-matched placebo cohorts, by horizon.}
    \label{fig:placebo-hist}
\end{figure}

\paragraph{Feature embargo.}

A second concern is mechanical rather than economic: the grid samples the last committed block,
so an increment could in principle reflect a hundred milliseconds of misalignment rather than
information. We test this directly by delaying \emph{every} feature, including both the anonymous
and identity blocks, by 200 and by 300 milliseconds while leaving the target at the grid stamp,
so that the endpoint of a one-second forecast lies 1.2 or 1.3 seconds after the newest
information. If misalignment explains the identity gain, these delays should make it collapse.
Instead,
Table~\ref{tab:embargo} shows a smooth decay under both learners: under ridge the one-second gain
falls from 13.2\% to 10.2\% to 9.1\%, with $t=8.2$ under both delays, and under trees from
6.0\% to 4.5\% to 4.0\%. This smooth decay is inconsistent with a mechanical misalignment
explanation. The delay reduces the anonymous benchmark far more than the identity increment. At
one second, the benchmark falls from 10.88\% to 6.40\% under ridge and from 19.48\% to 9.64\%
under trees. This disproportionate decline is consistent with the public order-book state
becoming stale faster than the identity increment.

\begin{table*}[t]\centering
\begin{threeparttable}
\caption{Feature embargo: identity gain when every feature is stale}
\label{tab:embargo}
\footnotesize\setlength{\tabcolsep}{4pt}
\begin{tabular}{lrrrrrrrrr}
\toprule
 & \multicolumn{3}{c}{No embargo} & \multicolumn{3}{c}{Features delayed 200\,ms}
 & \multicolumn{3}{c}{Features delayed 300\,ms} \\
\cmidrule(lr){2-4} \cmidrule(lr){5-7} \cmidrule(lr){8-10}
Horizon & Anonymous & + Identity & $t$ & Anonymous & + Identity & $t$
 & Anonymous & + Identity & $t$ \\
 & $R^2$ (\%) & Gain (\%) & & $R^2$ (\%) & Gain (\%) & & $R^2$ (\%) & Gain (\%) & \\
\midrule
\multicolumn{10}{l}{\textit{Panel A: Ridge (benchmark)}} \\
\midrule
0.2s & 6.91 & +18.7 & 8.1 & 4.74 & +14.6 & 8.3 & 3.49 & +12.2 & 8.1 \\
0.5s & 10.16 & +15.8 & 8.3 & 6.51 & +12.4 & 9.1 & 5.23 & +10.8 & 9.1 \\
1s & 10.88 & +13.2 & 9.2 & 7.61 & +10.2 & 8.2 & 6.40 & +9.1 & 8.2 \\
2s & 10.60 & +10.0 & 8.5 & 8.02 & +8.0 & 8.8 & 7.04 & +7.2 & 8.5 \\
5s & 9.17 & +6.7 & 9.5 & 7.55 & +5.5 & 9.0 & 6.89 & +5.1 & 7.5 \\
10s & 7.18 & +5.2 & 9.9 & 6.14 & +4.4 & 9.4 & 5.71 & +4.2 & 9.1 \\
30s & 3.56 & +3.0 & 6.1 & 3.14 & +2.6 & 5.4 & 2.96 & +2.4 & 5.3 \\
\midrule
\multicolumn{10}{l}{\textit{Panel B: Gradient-boosted trees}} \\
\midrule
0.2s & 19.26 & +10.7 & 5.4 & 8.85 & +10.3 & 4.6 & 6.00 & +6.2 & 4.8 \\
0.5s & 21.00 & +9.2 & 4.6 & 11.14 & +6.1 & 4.8 & 8.42 & +5.0 & 5.2 \\
1s & 19.48 & +6.0 & 5.0 & 11.97 & +4.5 & 5.2 & 9.64 & +4.0 & 6.0 \\
2s & 16.35 & +4.2 & 6.0 & 11.40 & +3.5 & 5.8 & 9.69 & +3.3 & 5.6 \\
5s & 12.00 & +3.0 & 6.1 & 9.44 & +2.7 & 6.1 & 8.48 & +2.3 & 4.1 \\
10s & 8.55 & +2.5 & 6.4 & 7.10 & +2.4 & 5.6 & 6.52 & +2.3 & 4.2 \\
30s & 3.74 & +2.7 & 1.6 & 3.26 & +3.8 & 3.5 & 3.07 & +3.1 & 1.6 \\
\bottomrule
\end{tabular}
\begin{tablenotes}[flushleft]\footnotesize
\item \textit{Notes:} The embargo delays every feature---the anonymous block and the identity
block alike---by 200\,ms or 300\,ms, while the target is left at the grid stamp, so the endpoint
of a one-second forecast lies 1.2 or 1.3 seconds after the newest information. Frozen wallet score,
top decile; sample, variable blocks, learners and $t$ construction are those of
Table~\ref{tab:part2-v1}. \emph{Anonymous} is the benchmark without identity features;
Gain $=R^2_{\mathrm{+Identity}}/R^2_{\mathrm{Anonymous}}-1$ is
the relative increase over that learner's own benchmark, and $t$ comes from the seven paired daily
MSE differences.
\end{tablenotes}
\end{threeparttable}\end{table*}

\subsection{Identity and Subsequent Order Flow}
\label{sec:policy}

The forecasts in Section~\ref{sec:prediction-results} are evaluated at every stamp on the
100-millisecond grid, and at most of those stamps nothing trades. Whether identity still helps at
the fills that actually occur is a separate question, and it is the one that matters for a
liquidity provider, since those are the moments at which a resting quote is taken. The increment is
larger there: 2.47 percentage points of one-second $R^2$ at an arrival, against 1.43 across the
grid as a whole. Forecasting is also easier at a fill, where the anonymous benchmark is about twice
as strong, so the comparison is one of explained return variance rather than of relative gain.

The simplest explanation of that larger increment is that it is a readout of order flow: seeing
historically toxic wallets resting at the quote indicates that a trade is about to arrive and in
which direction, and price follows the trade. Under this interpretation, the result reflects speed
rather than information in wallet histories, and the increment should disappear once order flow is
accounted for. We examine this explanation in two complementary ways. The first reveals the
direction of the focal arriving order; the second restricts the sample to fills followed by no
trade from another parent order.

Both exercises begin with the 3.09 million maker fills in the three model markets over July
21--27, each matched to the last grid observation at least 100 milliseconds before its consensus
fill time, so that every feature precedes the fill on the committed-tape clock. Models are fitted
on July 11--20 fills, and the arriving taker's wallet address is never a predictor.

\paragraph{Revealing the direction.} The first exercise adds the realized side $q=\pm1$ of the
focal order to both models and interacts it with every predictor, allowing predictor slopes to
differ between buyer- and seller-initiated fills. Because side is realized after the feature
timestamp, this is an ex post decomposition rather than a feasible forecast. At one second,
identity raises $R^2$ by 2.47 percentage points before side is observed ($t=10.8$) and by 1.50
points once side is revealed ($t=8.1$), a reduction of 39\%. The conditional increment becomes
statistically indistinguishable from zero at ten seconds (Table~\ref{tab:econ}).

\begin{table}[H]\centering
\begin{threeparttable}
\caption{Identity increment after revealing side and restricting subsequent trades}
\label{tab:econ}
\small\setlength{\tabcolsep}{7pt}\renewcommand{\arraystretch}{1.15}
\begin{tabular}{lrrrrrrr}
\toprule
 & \multicolumn{2}{c}{Pre-arrival} & \multicolumn{2}{c}{Realized side revealed}
 & \multicolumn{3}{c}{No other parent order} \\
\cmidrule(lr){2-3} \cmidrule(lr){4-5} \cmidrule(lr){6-8}
Horizon & $\Delta R^2$ (pp) & $t$ & $\Delta R^2$ (pp) & $t$ & $\Delta R^2$ (pp) & $t$
 & Fills (000) \\
\midrule
0.5s & +2.53 & 7.4 & +1.49 & 4.0 & +1.09 & 7.9 & 1,492 \\
1s & +2.47 & 10.8 & +1.50 & 8.1 & +1.04 & 5.5 & 817 \\
2s & +1.79 & 4.9 & +1.41 & 10.6 & +1.03 & 2.5 & 347 \\
5s & +1.47 & 3.5 & +1.26 & 5.9 & +0.49 & 0.9 & 86 \\
10s & +0.87 & 2.1 & +0.63 & 1.6 & +0.04 & 0.2 & 20 \\
30s & +0.45 & 1.8 & +0.35 & 1.4 & +0.04 & 0.5 & 2 \\
\bottomrule
\end{tabular}
\begin{tablenotes}[flushleft]\footnotesize
\item \textit{Notes:} Identity $\Delta R^2$ is the change in out-of-sample $R^2$ from adding
identity features to the anonymous benchmark for the forward midpoint return at the stated horizon.
Each of the 3.09 million evaluation-window maker fills is paired with the last nominal 100ms grid
timestamp at least 100ms before its consensus fill time, and inputs use the last commit available
by that timestamp.
\item \emph{Pre-arrival} uses only predictors observed before the focal fill and excludes its side.
\emph{Realized side revealed} adds $q=\pm1$ to both models and interacts it with every predictor;
this is an ex post decomposition. \emph{No other parent order} restricts both fit and evaluation
to fills followed by no trade from a different parent order within the horizon. Further fills from
the focal parent order are allowed, and this column does not condition on $q$. The subsample is
defined by a realized future property and shrinks sharply with the horizon, which is why its fill
count is reported. All $t$ statistics come from seven paired daily differences.
\end{tablenotes}
\end{threeparttable}\end{table}

Realized side therefore accounts for a meaningful share of the identity increment but does not
summarize it. Over the second beginning at the fill, and counting the arriving order itself, the
sign of net signed taker flow agrees with $q$ for 86.6\% of fills, while $q$ explains only 4.0\%
of that flow's variance. The exercise provides a strong control for
flow direction but leaves variation in order size, sweep intensity, and subsequent flow.

\paragraph{Restricting subsequent flow.} The side of the focal order need not summarize the orders
that follow it. After excluding the focal parent order, the sign of residual one-second flow agrees
with $q$ only 57.8\% of the time. We therefore repeat the prediction comparison among fills followed
by no trade from a different parent order within the forecast horizon. This restriction holds for
26.4\% of evaluation fills at one second and 48.3\% at half a second. Further fills from the same
parent order remain part of the focal event.

Within this restricted sample, identity raises $R^2$ by 1.04 percentage points at one second
($t=5.5$, 816{,}944 observations) and by 1.09 points at half a second ($t=7.9$, 1{,}491{,}847
observations). Because the restriction uses a realized future outcome, it is a mechanism
diagnostic rather than a feasible forecast. It also changes the market composition and becomes
increasingly selective at longer horizons, leaving 86{,}000 fills at five seconds and 20{,}000 at
ten seconds.

Together, the two exercises narrow an order-flow explanation without ruling it out. Identity's
predictive content is not summarized by the focal order's direction, and it remains among fills
followed by no other parent order. The tests do not distinguish anticipation of the focal order's
size and impact from subsequent quote revision. They show that wallet histories contain information
about the price response that is not captured by the anonymous market state or the focal order's
side alone. Translating this forecasting increment into implementable profits would require a
dynamic execution and market-making model accounting for latency, queue priority, fill
probabilities, fees, and inventory, which is beyond the scope of this paper;
Appendix~\ref{app:lean} instead reports the increment in payoff units.

\section{Conclusion}
\label{sec:conclusion}

Hyperliquid's public Level-4 record makes it possible to reconstruct the limit order book while
linking trading activity through persistent wallet addresses. Using this structure, we document
substantial persistence in address-level markouts: the rank correlation of wallet toxicity
across adjacent ten-day windows is $\rho=0.52$. This persistence also translates into return
predictability. In the three most active markets, adding features constructed from previously
scored toxic wallets raises one-second out-of-sample $R^2$ by 13.2\% relative to the prespecified
anonymous benchmark in a linear ridge model. Both findings recur in an independent December 2025
sample.

The analysis focuses on aggressive fills and quotes, but the same transparency opens several
directions for future research. Wallet-level cancellations and rejected orders may contain
information not captured by executed trades and resting quotes. More broadly, linking the full
sequence of maker and taker decisions provides a foundation for evaluating address-based trading
strategies in a dynamic market environment with latency, order choice, queue priority, fill risk,
fees, and inventory constraints.

\bibliographystyle{plainnat}
\bibliography{references}

\clearpage
\appendix
\numberwithin{table}{section}
\numberwithin{figure}{section}

\section{Details on the Data}
\label{app:reconstruction}

This appendix documents how the Level-4 record was collected and reconstructed, the
invariants the reconstruction enforces, and the diagnostics that make both auditable. The
guiding distinction is between \emph{coverage} (what was collected, established by block
counts, gaps, and archive hashes) and \emph{reconstruction validity} (whether an emitted book
state agrees with an independent contemporaneous oracle). Fill pairing and replay diagnostics
provide additional checks. We report each separately.

\subsection{Collection}
\label{app:collection}

A non-validating Hyperliquid node (\texttt{hl-node}, version pinned in the replication
artifact; managed by
the exchange's visor process) runs continuously on a dedicated cloud instance (8 vCPUs, 64~GB
memory), connects to the peer-to-peer network through bootstrap peers, and writes five raw
streams as they arrive: order statuses, raw book diffs, fills, TWAP statuses, and
miscellaneous events. Each stream is stored as hourly zstd-compressed NDJSON whose lines are
\emph{block envelopes} with fields \texttt{local\_time}, \texttt{block\_time},
\texttt{block\_number}, and \texttt{events}---so
the consensus block is the atomic unit of the archive, and record-level data are the exploded
\texttt{events} arrays. In parallel, the \texttt{metaAndAssetCtxs} API is sampled every ten
seconds for mark, oracle, and open-interest state. The July 1--30, 2026 collection window
contains two material infrastructure events. On July 27 the collector fell behind the chain from
13:30 UTC, peaking at thirty-five minutes of lag, and while that far behind it shed blocks
outright, leaving two holes in the tape of thirty-three and thirty-one minutes beginning at
15:05:23 and 17:05:34 UTC; it had recovered by 22:10 UTC. The oracle gate closed for all of the
first hole in all three markets and for the second's final seventeen minutes, so that time is
suppressed rather than emitted. During the opening fifteen minutes of the second hole the
reconstruction held a stale book that still agreed with the oracle to within tolerance---the
staleness limitation of a midpoint gate rather than a reconstruction error, and the reason we
report the gate as validating the midpoint and not the book. The collector then ran normally
until it failed outright at 06:06:45 UTC on July 28, and service resumed at 04:23:20 UTC on
July 29; from July 30 a
second, independently operated node provides the tape, spliced at 20:00 UTC (primary node
through hour 19, second node from hour 20). All analyses in the paper end on July 27; the
July 29--30 extension, where used, is stated explicitly.

\emph{Media damage and recovery.}\label{app:damage} Fourteen archived source files suffered disk
CRC read errors: eleven order-status hours on July 3--7 and July 21, and three book-diff hours on
July 20. GNU \texttt{ddrescue}
recovered them non-destructively: 18.36~GB of readable bytes were retained and 16.78~MB
remain unresolved across 1{,}337 mapped gap ranges, zero-filled in the recovered copies and
never assumed byte-complete. Nine book-diff files with verified deterministic decompression
failures are consumed only up to their probed last good block; their unreadable suffixes are
the only lost data, and each creates a replay seam. Together with the collection and node
boundaries, these seams produce thirteen reconstruction segments. Downstream, the reconstruction treats damage conservatively: any
unexpected decompression error aborts the segment (the whitelist covers exactly the nine
probed files), the book restarts \emph{empty} at the first intact post-gap file, and nothing
is emitted until a contemporaneous-oracle gate readmits the state---twelve consecutive
ten-second oracle ticks agreeing with the reconstructed mid within 25 basis points, with any
30-second tick gap closing the gate and resetting the streak. Later gate failures suppress
new state until the gate reopens. This is gate-revalidated BBO reconstruction, not exact full-book
recovery: latent resting-order errors can survive while the mid agrees, a limitation we
disclose rather than assume away.

\subsection{Reconstruction conventions and checks}
\label{app:reconstruction-checks}

Reconstruction proceeds in five deterministic steps, each chosen so that the output is a
pure function of the consensus-ordered block sequence.

\emph{(i) Segmented replay under a zero-corruption policy.} The book is rebuilt by replaying
order-status and book-diff events block by block. Each segment starts from an empty book; no
state is emitted until the oracle gate admits the reconstruction. An unexpected decompression
or parse error aborts the segment rather than skipping records; each of the nine verified
damaged files is consumed only through its last good block. \emph{(ii) Block-close
state.} A midpoint or depth quantity evaluated at any instant $t$ is the block-close state of
the last block committed at or before $t$---a right-continuous, backward-looking step
function. \emph{(iii) Fill pairing.} Every execution produces one record per side;
pairing on (day, market, trade identifier) recovers maker and taker for each fill.
\emph{(iv) Aggressive-order grouping.} Fills sharing (wallet, order identifier, market) are
one aggressive order---one trading decision. \emph{(v) Lifecycle linkage.} Order-status
events link placement, modification, cancellation, and execution by order identifier within
the collected stream.

Replay is applied in completed-block order. A block-number rewind or crossed book triggers a
book reset and closes the gate; aggregate depth is kept non-negative. Reconstructed state is
emitted only while its midpoint agrees with the independent oracle. This gate validates the
midpoint, not every latent resting order or the full depth profile, so full-book errors can
survive while the midpoint agrees.

Two identity variables are shares of book depth---the toxic decile's weight at the touch and its
signed contribution to the touch imbalance---and are therefore bounded by one in absolute value
by construction. On a handful of coin-days the top of book prints momentarily near-empty, and a
ratio computed against that denominator can exceed its own bound. We clamp the denominator to be
at least the numerator, which enforces the bound at every observation, and the feature builder
asserts it before writing. Five coin-days inside the sample are affected, and all reported
results use the corrected features.

Gate closures are rare, and where they occur they are shared across markets. Over July 1--27 the
gate is closed for 0.97\% of coin-time in the three sample markets, rising to 1.83\% inside the
July 21--27 evaluation window. Across the full sample the median closure lasts 123
seconds---the twelve-tick
re-admission streak itself, so the typical closure is a brief re-warmup after one disagreeing
tick---while suppressed time is dominated by a short tail of long episodes, the longest running
70 minutes. Those episodes are market-wide rather than coin-specific: 81\% of suppressed time in
the three sample markets is common to all three at once, which is what an oracle-side outage
looks like and not what selective reconstruction failure would look like. The distinction
matters, because state that went missing precisely when one book was hard to rebuild would bias
the sample, whereas losing the same clock window in every market does not. The replay also logs
its own discontinuities---eight node-restart rewinds across the production segments, each
forcing a book reset and gate re-admission, and 132 crossed-book self-heals---and a second
reconstruction of the same raw stream, run separately, reproduces both counts segment by
segment.

That the segment boundaries themselves are harmless can be checked directly. July 12 lies deep
inside a continuous six-day segment, so its book was built by a replay that had been running for
two days. Replaying July 12 on its own---empty book, warm-up, gate re-admission---reproduces the
continuous replay's bid, ask, mid, and spread on 99.994\% of the 1.32 million comparable blocks
in the three sample markets; the exceptions are 78 blocks on one coin inside the first thirty-one
minutes, worth at most 0.8 basis points of the mid. The gate opened 115 seconds after the empty-book
restart, hours before the production window began, so the restart cost no production time.

The downstream screens differ by exercise. Markout construction requires fresh pre-event and
endpoint states and sets gate-suppressed endpoints to missing. The forecasting grid truncates
targets at UTC-day boundaries but carries the last emitted block state forward and does not
apply the markout-specific staleness screens. This makes collection continuity especially
important for the forecasting sample; Table~\ref{tab:audit} exposes every gap in that window.

\subsection{Audit}
\label{app:audit}

The audit comprises three components. \emph{Fill pairing:} of 27{,}902{,}380 crossed (taker) fill
records on July 1--27, every one pairs with a maker-side record at the same (day, market,
trade identifier)---zero unmatched legs, by count and by notional. \emph{Archive
integrity:} a SHA-256 manifest covers every archived hourly shard of all six streams---the
five node streams and the oracle sample (180
day-stream shard sets). \emph{Collection continuity:} a forecasting result could be distorted if
the archive selectively missed active periods. The day-by-day block census in
Table~\ref{tab:audit} weighs against this concern. Twenty-two of the thirty
collection days show a maximum intra-day block gap of three or fewer with no gap of ten
blocks or more. Every fitting-window day (July 11--20) satisfies that block-gap criterion, as does
the evaluation window apart from a sixteen-block gap on July 25 and a 28{,}326-block gap on
July 27. The remaining interruptions are documented: larger block gaps in the scoring window
(July 1--2), the truncated shards described above, and the July 28--29 node
outage, which appears as a 1{,}119{,}241-block discontinuity across that boundary. July 30 is
continuous once the second node's hours are merged, so the 20:00 UTC splice introduces no
break in the block sequence.

\begin{table}[H]\centering
\begin{threeparttable}
\caption{Collection audit by day: block-level continuity}
\label{tab:audit}
\footnotesize\setlength{\tabcolsep}{6pt}
\begin{tabular}{lcrrrrrr}
\toprule
Day & Hours & Blocks (m) & Max gap & $\geq$10 & Absent (\%) & Day gap & Trunc. \\
\midrule
\multicolumn{8}{l}{\textit{Scoring window: July 1--10}} \\
\addlinespace[1pt]
7/1 & 24 & 1.150 & 27,900 & 1 & 8.3 & --- & 0 \\
7/2 & 24 & 1.097 & 59,415 & 3 & 12.8 & -1 & 1 \\
7/3 & 24 & 1.184 & 4 & 0 & 5.8 & -1 & 2 \\
7/4 & 24 & 1.151 & 7 & 0 & 5.1 & 0 & 5 \\
7/5 & 24 & 1.128 & 3 & 0 & 5.3 & 0 & 3 \\
7/6 & 24 & 1.132 & 4 & 0 & 5.7 & -1 & 3 \\
7/7 & 24 & 1.138 & 3 & 0 & 5.8 & 0 & 1 \\
7/8 & 24 & 1.147 & 3 & 0 & 5.5 & 0 & 1 \\
7/9 & 24 & 1.158 & 3 & 0 & 5.2 & 1 & 1 \\
7/10 & 24 & 1.154 & 3 & 0 & 5.2 & 1 & 1 \\
\midrule
\multicolumn{8}{l}{\textit{Fitting window: July 11--20}} \\
\addlinespace[1pt]
7/11 & 24 & 1.138 & 3 & 0 & 5.1 & -1 & 0 \\
7/12 & 24 & 1.148 & 3 & 0 & 5.2 & 0 & 0 \\
7/13 & 24 & 1.148 & 3 & 0 & 5.8 & 0 & 0 \\
7/14 & 24 & 1.147 & 3 & 0 & 5.6 & 0 & 0 \\
7/15 & 24 & 1.142 & 3 & 0 & 5.6 & 0 & 1 \\
7/16 & 24 & 1.146 & 3 & 0 & 5.4 & 0 & 2 \\
7/17 & 24 & 1.141 & 3 & 0 & 5.7 & 0 & 0 \\
7/18 & 24 & 1.147 & 3 & 0 & 5.3 & -1 & 0 \\
7/19 & 24 & 1.151 & 2 & 0 & 5.2 & 0 & 0 \\
7/20 & 24 & 1.149 & 3 & 0 & 5.3 & 0 & 3 \\
\midrule
\multicolumn{8}{l}{\textit{Evaluation window: July 21--27}} \\
\addlinespace[1pt]
7/21 & 24 & 1.149 & 3 & 0 & 5.2 & 0 & 4 \\
7/22 & 24 & 1.148 & 2 & 0 & 5.0 & -1 & 0 \\
7/23 & 24 & 1.147 & 3 & 0 & 5.0 & 0 & 0 \\
7/24 & 24 & 1.151 & 3 & 0 & 4.9 & 0 & 0 \\
7/25 & 24 & 1.152 & 16 & 1 & 4.6 & 1 & 0 \\
7/26 & 24 & 1.152 & 3 & 0 & 4.8 & 0 & 0 \\
7/27 & 24 & 1.100 & 28,326 & 2 & 9.4 & 1 & 0 \\
\midrule
\multicolumn{8}{l}{\textit{Post-sample: July 28--30}} \\
\addlinespace[1pt]
7/28 & 7 & 0.293 & 2 & 0 & 4.9 & -1 & 0 \\
7/29 & 20 & 0.871 & 22,008 & 2 & 8.5 & 1,119,241 & 0 \\
7/30 & 21+10 & 1.177 & 3 & 0 & 4.8 & 0 & 0 \\
\bottomrule
\end{tabular}
\begin{tablenotes}[flushleft]\footnotesize
\item \textit{Notes:} Census of distinct block numbers in the archived order-status and
book-diff envelopes of both collection nodes, grouped by the window each day belongs to. July 28
is in the post-sample block but is excluded from the sample by the collection outage.
\item Hours counts hourly shards carrying at least one block, primary node plus (after $+$) the
second node. Max gap is the largest jump between adjacent observed block numbers within the day;
$\geq$10 counts jumps of at least ten blocks. Absent is the share of block heights missing from
the day's observed-height span. Day gap reports missing heights between adjacent daily spans;
$-1$ denotes a duplicated boundary height and 0 denotes contiguity. Trunc. counts source shards
decoded only through their verified last good record.
\item All fitting-window days have maximum gaps of three blocks or fewer; the evaluation-window
gap exceptions are July 25 (16 blocks) and July 27 (28{,}326 blocks). July 30 is continuous after
merging the second node from 20:00 UTC.
\end{tablenotes}
\end{threeparttable}
\end{table}

\section{Additional Results and Robustness}
\label{app:altspec}

\subsection{Peer-Adjusted Wallet Score}
\label{app:peer-score}

The main text defines wallet toxicity by the raw ten-second markout. This subsection repeats the
toxicity validation and forecasting race with a peer-adjusted score, which measures how much more a wallet's
orders mark out than the orders of \emph{other} wallets trading the same coin in the same
minute. The peer benchmark removes anything common to the market at that instant, so the
comparison answers the natural objection that the identity block is picking up moments when
everyone does well rather than wallets who do well.

\begin{figure}[H]
    \centering
    \begin{minipage}{0.82\textwidth}
        \centering
        \includegraphics[width=\textwidth]{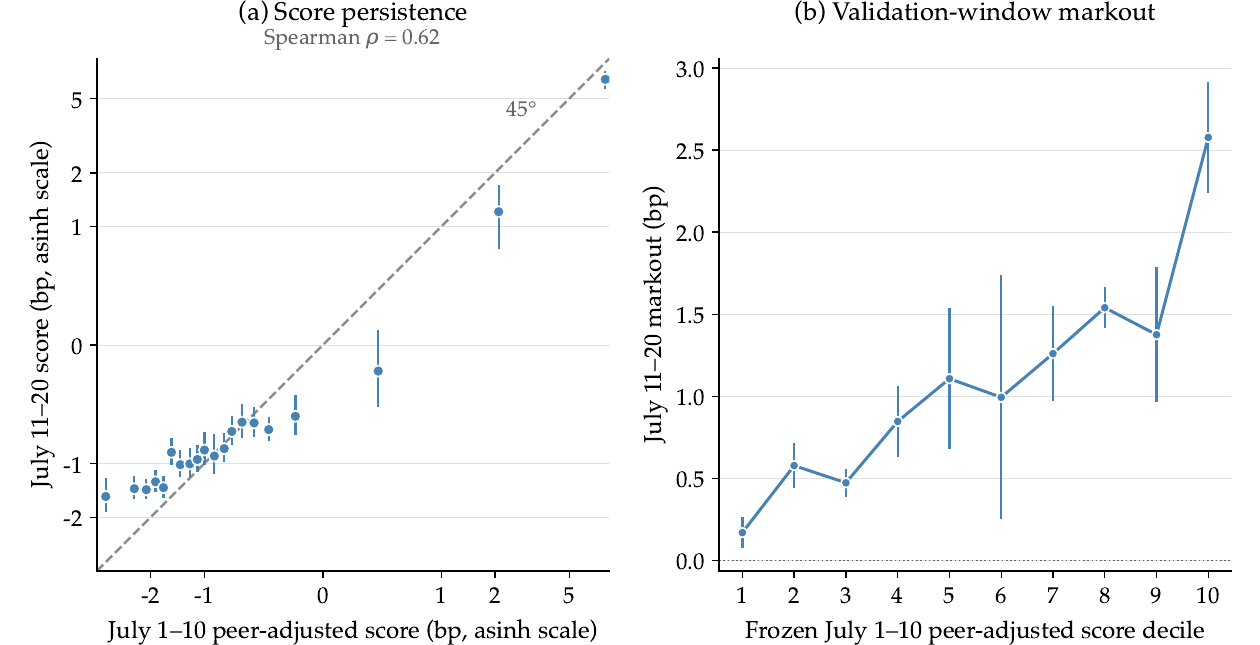}
        \caption{Wallet toxicity using the peer-adjusted score. Panel (a) shows score persistence; Panel (b) shows validation-window markouts by frozen-score decile. Bars are 95\% confidence intervals.}
        \label{fig:persistence-cont}
    \end{minipage}
\end{figure}

\paragraph{The ranking survives peer adjustment, and persists better.}
If the raw ranking merely captures market-wide timing, removing the same-coin, same-minute
component should weaken its persistence. Instead, the peer-adjusted rank has a Spearman
$\rho$ of 0.62 across adjacent ten-day windows, above the raw score's 0.52
(Figure~\ref{fig:persistence-cont}(a)).

The peer-adjusted score also retains the economically relevant upper tail: validation markouts
rise sharply across the top deciles (Figure~\ref{fig:persistence-cont}(b)). The accompanying
decile portrait in Table~\ref{tab:part1-panels-cont} further shows that the top decile pays the
lowest median taker fee.

\begin{table}[H]\centering
\begin{threeparttable}
\caption{Wallet deciles under the peer-adjusted score}
\label{tab:part1-panels-cont}
\footnotesize\setlength{\tabcolsep}{5pt}
\begin{tabular}{lrrrrr}
\toprule
 & \multicolumn{4}{c}{Scoring window (July 1--10)} & \multicolumn{1}{c}{Validation window (July 11--20)} \\
\cmidrule(lr){2-5} \cmidrule(lr){6-6}
 & Wallets & Orders & Median & Median fee & Markout \\
Decile & & & order (\$) & (bps) & (bps) \\
\midrule
D1 & 206 & 56,546 & 1,852 & 4.50 & 0.17 \\
D2 & 206 & 77,396 & 1,747 & 4.50 & 0.58 \\
D3 & 206 & 62,668 & 1,078 & 4.50 & 0.47 \\
D4 & 207 & 118,149 & 578 & 4.50 & 0.85 \\
D5 & 206 & 44,405 & 177 & 4.50 & 1.11 \\
D6 & 206 & 55,167 & 100 & 4.50 & 0.99 \\
D7 & 207 & 115,782 & 82 & 4.50 & 1.26 \\
D8 & 206 & 178,964 & 58 & 4.50 & 1.54 \\
D9 & 206 & 67,004 & 50 & 4.50 & 1.38 \\
\midrule
D10 & 206 & 242,721 & 1,007 & 2.97 & 2.58 \\
\bottomrule
\end{tabular}
\begin{tablenotes}[flushleft]\footnotesize
\item \textit{Notes:} Deciles use the July 1--10 peer-adjusted score and are never
re-sorted; the peer benchmark is the notional-weighted mean markout of other wallets' orders in
the same coin and minute, so the score measures excess markout rather than level.
\item Wallets, orders, median order, and median fee are measured in the scoring window; markout
is the notional-weighted ten-second markout measured out of sample during July 11--20, which is
the only statistic the peer-adjusted arm carries for the validation group. D10, the toxic
decile, is set off by a rule.
\end{tablenotes}
\end{threeparttable}\end{table}

\paragraph{The predictive gain survives, at roughly half the size.}
If common market timing drives the forecasting result, peer adjustment should also eliminate the
identity gain. Against the identical public benchmark, Table~\ref{tab:part2-cont} shows that the
short-horizon gain remains positive and precisely estimated: at one second it
is 6.8\% for the top decile, roughly half the raw score's 13.2\%. Because peer adjustment
removes the same-coin, same-minute component by construction, the remaining gain shows that
the identity block is not solely a proxy for common market timing.

\begin{table}[H]\centering
\begin{threeparttable}
\caption{Return prediction under the peer-adjusted score}
\label{tab:part2-cont}
\footnotesize\setlength{\tabcolsep}{5pt}
\begin{tabular}{lrrrrrr}
\toprule
 & \multicolumn{3}{c}{Ridge (benchmark)} & \multicolumn{3}{c}{Gradient-boosted trees} \\
\cmidrule(lr){2-4} \cmidrule(lr){5-7}
Horizon & + Identity & Gain (\%) & $t$ & + Identity & Gain (\%) & $t$ \\
 & $R^2$ (\%) & & & $R^2$ (\%) & & \\
\midrule
\multicolumn{7}{l}{\textit{Panel: Frozen score, top decile}} \\
\midrule
0.2s & 7.46 & +7.9 & 5.6 & 20.69 & +7.4 & 5.5 \\
0.5s & 10.86 & +6.9 & 5.0 & 22.28 & +6.1 & 6.6 \\
1s & 11.62 & +6.8 & 5.1 & 20.27 & +4.1 & 5.6 \\
2s & 11.23 & +5.9 & 4.2 & 16.91 & +3.4 & 6.2 \\
5s & 9.62 & +4.9 & 4.2 & 12.33 & +2.8 & 5.7 \\
10s & 7.49 & +4.3 & 4.4 & 8.72 & +2.0 & 5.0 \\
30s & 3.67 & +3.3 & 4.1 & 3.84 & +2.7 & 3.9 \\
\midrule
\multicolumn{7}{l}{\textit{Panel: Frozen score, top quintile}} \\
\midrule
0.2s & 7.57 & +9.6 & 5.5 & 20.94 & +8.7 & 5.3 \\
0.5s & 11.04 & +8.7 & 5.3 & 22.48 & +7.1 & 5.5 \\
1s & 11.78 & +8.3 & 5.4 & 20.35 & +4.5 & 4.8 \\
2s & 11.34 & +7.0 & 4.6 & 16.92 & +3.5 & 5.2 \\
5s & 9.68 & +5.5 & 4.6 & 12.31 & +2.6 & 5.8 \\
10s & 7.53 & +4.8 & 4.7 & 8.77 & +2.6 & 4.9 \\
30s & 3.69 & +3.8 & 4.4 & 3.83 & +2.4 & 2.1 \\
\midrule
\multicolumn{7}{l}{\textit{Panel: Rolling ten-day score, top decile}} \\
\midrule
0.2s & 7.65 & +10.7 & 4.2 & 20.05 & +4.1 & 1.8 \\
0.5s & 11.05 & +8.8 & 3.8 & 21.54 & +2.6 & 0.8 \\
1s & 11.74 & +7.9 & 4.3 & 20.02 & +2.8 & 1.8 \\
2s & 11.25 & +6.2 & 4.1 & 16.83 & +3.0 & 4.2 \\
5s & 9.59 & +4.6 & 4.8 & 12.27 & +2.3 & 3.8 \\
10s & 7.46 & +3.9 & 4.5 & 8.71 & +1.9 & 3.1 \\
30s & 3.66 & +2.8 & 3.1 & 3.86 & +3.2 & 3.1 \\
\midrule
\multicolumn{7}{l}{\textit{Panel: Rolling ten-day score, top quintile}} \\
\midrule
0.2s & 7.87 & +13.9 & 5.5 & 21.20 & +10.1 & 3.4 \\
0.5s & 11.36 & +11.8 & 5.2 & 22.72 & +8.2 & 4.7 \\
1s & 11.98 & +10.1 & 5.4 & 20.52 & +5.3 & 5.1 \\
2s & 11.41 & +7.6 & 4.8 & 17.02 & +4.1 & 5.5 \\
5s & 9.65 & +5.2 & 4.7 & 12.37 & +3.1 & 4.8 \\
10s & 7.49 & +4.3 & 4.6 & 8.73 & +2.2 & 4.1 \\
30s & 3.66 & +2.9 & 3.3 & 3.81 & +1.8 & 2.1 \\
\bottomrule
\end{tabular}
\begin{tablenotes}[flushleft]\footnotesize
\item \textit{Notes:} Table~\ref{tab:part2-v1} re-run with the wallet cohort defined by
the peer-adjusted score instead of the raw score, over the full grid of score definitions and
tail cuts. Scores are estimated July 1--10, models fit July 11--20, and evaluation is July
21--27. The \emph{+ Identity} columns report out-of-sample $R^2$, Gain
$=R^2_{\mathrm{+Identity}}/R^2_{\mathrm{Anonymous}}-1$ is the relative increase, and $t$
uses seven paired daily MSE differences.
\item The anonymous benchmark does not depend on how the cohort is selected, so it is the one
printed in Table~\ref{tab:part2-v1} and is not repeated here.
\end{tablenotes}
\end{threeparttable}\end{table}

\subsection{Cross-Venue Latency}
\label{app:latency}

Table~\ref{tab:latency} reports the diagnostic summarised in Section~\ref{sec:toxic-results},
decile by decile: how often each decile's orders arrive on the side of the oracle--midpoint gap,
the mean signed gap they trade into, and the markout that survives in the subsample where the gap
is at most one basis point.

\begin{table}[tbp]\centering
\begin{threeparttable}
\caption{Cross-venue latency diagnostic by wallet decile}
\label{tab:latency}
\footnotesize\setlength{\tabcolsep}{5pt}
\begin{tabular}{lrrrrr}
\toprule
 & \multicolumn{3}{c}{All orders} & \multicolumn{2}{c}{Gap $\leq 1$\,bp} \\
\cmidrule(lr){2-4} \cmidrule(lr){5-6}
Decile & Orders & Aligned (\%) & Mean gap (bp) & Orders & Markout (bp) \\
\midrule
D1 & 36,399 & 57.9 & -0.62 & 6,815 & +0.16 \\
D2 & 57,184 & 56.0 & -0.26 & 13,344 & +0.10 \\
D3 & 79,017 & 40.7 & -0.64 & 16,959 & -0.19 \\
D4 & 61,822 & 45.4 & -0.38 & 16,330 & -0.03 \\
D5 & 45,888 & 50.9 & -0.86 & 10,000 & +0.11 \\
D6 & 85,933 & 45.0 & -0.76 & 24,129 & +0.11 \\
D7 & 40,285 & 44.1 & -1.30 & 7,861 & +0.11 \\
D8 & 36,758 & 46.8 & -1.72 & 8,172 & +0.63 \\
D9 & 180,508 & 53.2 & -0.77 & 40,922 & +0.95 \\
D10 & 408,053 & 41.3 & -1.45 & 78,918 & +2.12 \\
\bottomrule
\end{tabular}
\begin{tablenotes}[flushleft]\footnotesize
\item \textit{Notes:} The venue's oracle price aggregates external venues and is sampled every ten seconds; the gap is that oracle price less the local midpoint immediately before each aggressive order. \emph{Aligned} counts orders arriving on the gap's side, which is what latency arbitrage would produce: a wallet reacting to an external move buys when the oracle sits above the local mid. \emph{Mean gap traded into} is the signed gap in basis points, negative when the order trades against the displacement. The final two columns restrict to orders facing a gap of at most one basis point, where there is nothing to arbitrage; these are 22\% of all orders. Fama--MacBeth daily D10$-$D1 markout coefficients over July 11--20 average 1.72 without gap controls and 1.75 with them. Because the oracle can be up to ten seconds stale, the test bounds a slow cross-venue channel but cannot rule out a subsecond one.
\end{tablenotes}
\end{threeparttable}\end{table}

\subsection{Tail Cut and Score Definition}
\label{app:score-alternatives}

The forecasting result would be fragile if it depended on the exact tail cut or on freezing the
score after July 10. Table~\ref{tab:part2-v1-appx} replaces the top decile with the top quintile
and the frozen score with a rolling ten-day score. Neither design choice materially changes the
gain at any reported horizon.

\begin{table}[H]\centering
\begin{threeparttable}
\caption{Identity gain by tail cut and score definition}
\label{tab:part2-v1-appx}
\footnotesize\setlength{\tabcolsep}{5pt}
\begin{tabular}{lrrrrrr}
\toprule
 & \multicolumn{3}{c}{Ridge (benchmark)} & \multicolumn{3}{c}{Gradient-boosted trees} \\
\cmidrule(lr){2-4} \cmidrule(lr){5-7}
Horizon & + Identity & Gain (\%) & $t$ & + Identity & Gain (\%) & $t$ \\
 & $R^2$ (\%) & & & $R^2$ (\%) & & \\
\midrule
\multicolumn{7}{l}{\textit{Panel: Frozen score, top quintile}} \\
\midrule
0.2s & 8.15 & +17.9 & 7.6 & 21.57 & +12.0 & 5.6 \\
0.5s & 11.73 & +15.5 & 7.7 & 23.10 & +10.0 & 5.2 \\
1s & 12.30 & +13.1 & 8.4 & 20.78 & +6.7 & 5.2 \\
2s & 11.68 & +10.2 & 8.8 & 17.09 & +4.6 & 5.7 \\
5s & 9.82 & +7.1 & 8.4 & 12.40 & +3.4 & 5.0 \\
10s & 7.57 & +5.5 & 11.2 & 8.79 & +2.8 & 4.8 \\
30s & 3.67 & +3.2 & 5.1 & 3.92 & +4.7 & 4.4 \\
\midrule
\multicolumn{7}{l}{\textit{Panel: Rolling ten-day score, top decile}} \\
\midrule
0.2s & 8.13 & +17.7 & 6.0 & 20.94 & +8.7 & 3.9 \\
0.5s & 11.69 & +15.1 & 5.9 & 22.71 & +8.1 & 4.1 \\
1s & 12.26 & +12.7 & 6.5 & 20.65 & +6.0 & 5.0 \\
2s & 11.61 & +9.6 & 6.0 & 17.11 & +4.7 & 5.9 \\
5s & 9.76 & +6.3 & 6.2 & 12.37 & +3.2 & 5.5 \\
10s & 7.52 & +4.8 & 6.9 & 8.77 & +2.6 & 5.2 \\
30s & 3.64 & +2.3 & 6.0 & 3.88 & +3.6 & 3.9 \\
\midrule
\multicolumn{7}{l}{\textit{Panel: Rolling ten-day score, top quintile}} \\
\midrule
0.2s & 8.07 & +16.7 & 5.9 & 21.36 & +10.9 & 5.6 \\
0.5s & 11.62 & +14.4 & 5.8 & 22.97 & +9.4 & 5.2 \\
1s & 12.20 & +12.2 & 6.3 & 20.70 & +6.3 & 5.4 \\
2s & 11.59 & +9.4 & 6.5 & 17.12 & +4.7 & 5.5 \\
5s & 9.76 & +6.4 & 6.4 & 12.40 & +3.3 & 5.1 \\
10s & 7.54 & +4.9 & 7.9 & 8.79 & +2.9 & 5.0 \\
30s & 3.65 & +2.6 & 4.3 & 3.86 & +3.2 & 2.9 \\
\bottomrule
\end{tabular}
\begin{tablenotes}[flushleft]\footnotesize
\item \textit{Notes:} Alternatives to the headline specification of
Table~\ref{tab:part2-v1}, which reports the frozen score at the top decile. Sample, targets,
variable blocks, learners and $t$ construction are identical, so the anonymous benchmark is
common to every panel and is the one printed in Table~\ref{tab:part2-v1} rather than repeated
here. The \emph{+ Identity} columns report out-of-sample $R^2$ and Gain is the
relative increase over that common benchmark.
\end{tablenotes}
\end{threeparttable}\end{table}

\subsection{Matched-Wallet Placebo}
\label{app:matched-placebo}

If size and activity alone explain the July result, a matched cohort should reproduce the toxic
cohort's forecasting gain. No matched cohort reproduces the gain through ten seconds. At one
second, the real top-decile $R^2$ increment is 1.44 percentage points
against 0.88 for the best of 200 activity-matched draws, so the best matched cohort reaches
61\% of the real increment. The rank
$p$ is $1/201$ at every horizon through ten seconds (Table~\ref{tab:p2placebo}). At thirty seconds the decile's increment of
0.11 percentage points falls inside a placebo range reaching 0.14 ($p=0.16$), while the top
quintile still exceeds every draw; the quintile is the wider cohort and averages over more
wallets, which is what keeps its increment above the ceiling once the signal is small. Exact-cell
donor limits make each draw smaller than the full toxic cohort, so the rank is descriptive rather
than an equal-size randomization test.

\begin{table}[H]\centering
\begin{threeparttable}
\caption{Matched-wallet placebo: the toxicity ranking against 200 activity-matched
cohorts}
\label{tab:p2placebo}
\small\setlength{\tabcolsep}{4.5pt}
\begin{tabular}{lrrrrrrr}
\toprule
 & & Real & \multicolumn{3}{c}{200 matched cohorts, $\Delta R^2$ (pp)} & & \\
\cmidrule(lr){4-6}
Horizon & Anonymous $R^2$ (\%) & $\Delta R^2$ (pp) & mean & p95 & max & max/real & $p$ \\
\midrule
\multicolumn{8}{l}{\textit{Top decile}} \\
\midrule
0.2s & 6.91 & +1.291 & +0.389 & +0.489 & +0.579 & 45\% & 0.0050 \\
0.5s & 10.16 & +1.601 & +0.596 & +0.741 & +0.851 & 53\% & 0.0050 \\
1s & 10.88 & +1.435 & +0.642 & +0.773 & +0.877 & 61\% & 0.0050 \\
2s & 10.60 & +1.062 & +0.534 & +0.651 & +0.690 & 65\% & 0.0050 \\
5s & 9.17 & +0.617 & +0.355 & +0.443 & +0.471 & 76\% & 0.0050 \\
10s & 7.18 & +0.374 & +0.229 & +0.296 & +0.308 & 82\% & 0.0050 \\
30s & 3.56 & +0.107 & +0.083 & +0.124 & +0.136 & 127\% & 0.1642 \\
\midrule
\multicolumn{8}{l}{\textit{Top quintile}} \\
\midrule
0.2s & 6.91 & +1.239 & +0.199 & +0.301 & +0.313 & 25\% & 0.0050 \\
0.5s & 10.16 & +1.574 & +0.289 & +0.435 & +0.451 & 29\% & 0.0050 \\
1s & 10.88 & +1.426 & +0.295 & +0.383 & +0.398 & 28\% & 0.0050 \\
2s & 10.60 & +1.078 & +0.215 & +0.305 & +0.317 & 29\% & 0.0050 \\
5s & 9.17 & +0.647 & +0.157 & +0.221 & +0.229 & 35\% & 0.0050 \\
10s & 7.18 & +0.392 & +0.106 & +0.147 & +0.157 & 40\% & 0.0050 \\
30s & 3.56 & +0.113 & +0.030 & +0.053 & +0.061 & 54\% & 0.0050 \\
\bottomrule
\end{tabular}
\begin{tablenotes}[flushleft]\footnotesize
\item \textit{Notes:} Panels give the identity block built from the top decile and the top
quintile of the frozen toxicity ranking. Cohort quote features incorporate every block-close update before being carried to the
100ms grid, matching the headline feature construction.
\item Draws match the real cohort cell by cell on deciles of
scoring-window notional crossed with order count, sampled without replacement; cells where the
cohort exhausts the donor pool are filled as far as they go, so matched cohorts are slightly
less extreme on activity than the real one.
\item Ridge benchmark, fit July 11--20 with
expanding-day validation, evaluated July 21--27; the placebo mean, p95 and max summarize the same
increment across the 200 draws. $p=(1+\#\{\mathrm{placebo}\ge
\mathrm{real}\})/201$, so $p=0.0050$ is the floor attained when no draw reaches the real
increment, and the max/real column reports the largest draw as a share of that increment.
Because some cells exhaust their donor pool, the rank
describes these generated cohorts rather than an exact equal-size randomization test.
\end{tablenotes}
\end{threeparttable}\end{table}

\section{Replication on December 2025}
\label{app:dec}

The core design is repeated on an independent public dataset: the December 2025 release of
Hyperliquid \textsc{btc}, \textsc{eth}, and \textsc{sol}, scored December 1--10, fit December
11--20, and evaluated December 21--31. The score definitions, predictor blocks, learners, and
horizons are unchanged. We report the persistence, forecasting, and matched-cohort exercises;
the order-arrival timing test is specific to July.
A useful replication should expose the design to different market conditions rather than simply
extend the original sample. The December sample provides this variation. Relative to July,
Table~\ref{tab:data-summary-dec2025} documents a deeper record of a narrower market set: three markets over
a full month carry more messages and more taker notional than July's ten over 27 days, on fewer
wallets and fewer aggressive orders. Spreads are wider throughout, so the replication is not a
re-run under easier conditions.

The public release differs from the July node stream in one respect that matters for
reconstruction: it does not retain the consensus-block envelope or the oracle snapshots, so the
book diffs carry no timestamp or block number. We recover timing by matching each diff on its
order identifier to the order-status and trade streams, which are timestamped: a new order takes
the time of its open status, an update the time of the trade that filled it, and a removal the
time of its cancel or final fill; about 99.98\% of diffs receive an exact time. File order is
kept as the sequence, the few anchors inconsistent with it are demoted, and any interval in which
the residual uncertainty exceeds 100~ms is closed and excluded. Because there is no oracle
stream, the December gate is internal, opening afresh after each closure or crossed-book reset
with a fixed warm-up, and quotes older than 120 seconds are screened as in July; a ``block'' in
Table~\ref{tab:data-summary-dec2025} is the group of diffs sharing one recovered time within a
gate, applied atomically.\footnote{The reconstruction code and the full audit tables (anchor coverage, closures,
replay integrity, and the trade-location and status-only-book cross-checks) are available at
the GitHub repository \url{https://github.com/daojingzhai/public-trader-identity}.}

\begin{table}[!htbp]
\centering
\footnotesize
\begin{threeparttable}
\caption{Level-4 sample by perpetual market, December 1--31, 2025}
\label{tab:data-summary-dec2025}
\setlength{\tabcolsep}{3.5pt}
\renewcommand{\arraystretch}{0.95}
\begin{tabular}{lrrrrrrr}
\toprule
Market & L4 messages & BBO updates & Agg. orders & Taker fills & Wallets & Notional & Spread \\
 & (bn) & (m) & (m) & (m) & (thousands) & (\$bn) & (bps) \\
\midrule
BTC & 15.08 & 1.91 & 4.28 & 11.92 & 64.7 & 75.1 & 0.38 \\
ETH & 7.82 & 1.83 & 3.62 & 6.84 & 40.2 & 42.6 & 0.84 \\
SOL & 3.35 & 1.58 & 1.62 & 3.10 & 34.8 & 9.8 & 1.08 \\
\midrule
All & 26.25 & 5.32 & 9.53 & 21.86 & 89.8 & 127.5 & 0.68 \\
\bottomrule
\end{tabular}
\begin{tablenotes}[flushleft]\footnotesize
\item \textit{Notes:} L4 messages combine raw order-status and book-diff records; BBO updates are gate-valid blocks in which the best quote changed. Aggressive orders, wallets, notional and spread come from the aggressive-event tape and taker fills are taker legs; spread is the mean quoted bid--ask spread immediately before aggressive orders. Wallets are distinct within market and deduplicated across markets in the All row.
\item The sample covers the three markets of the public December 2025 release, which carries accepted and rejected order statuses in separate archives; both are counted, matching the July node stream.
\end{tablenotes}
\end{threeparttable}
\end{table}

\begin{figure}[H]
    \centering
    \begin{minipage}{0.82\textwidth}
        \centering
        \includegraphics[width=\textwidth]{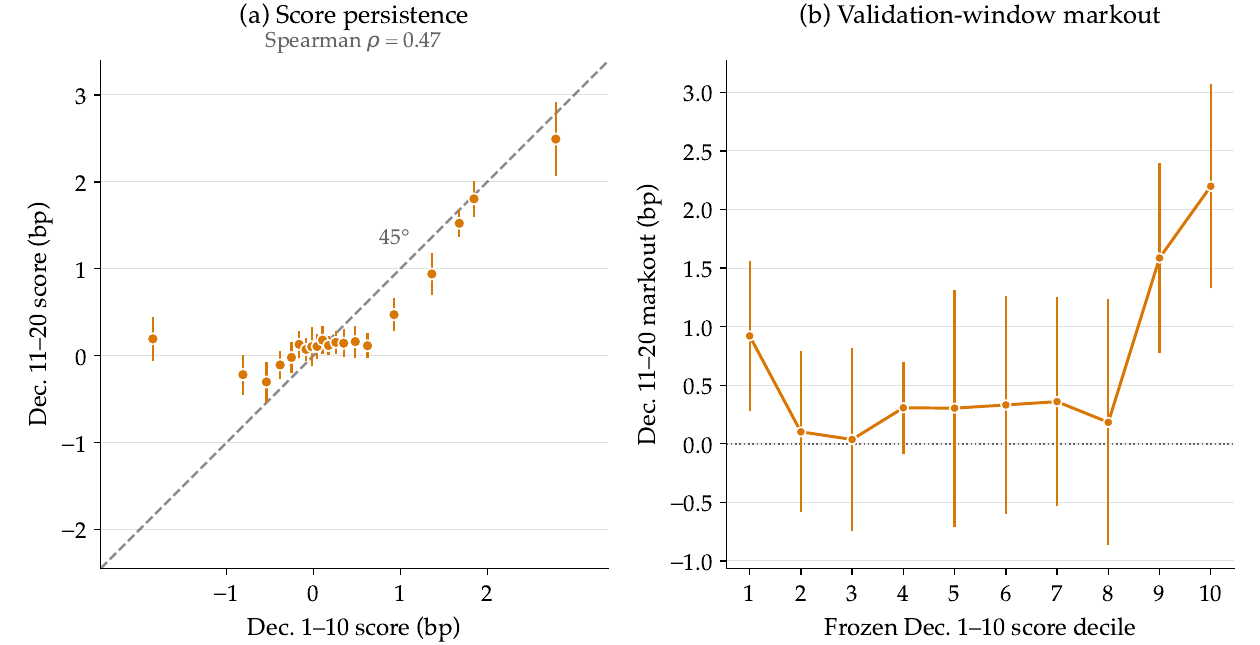}
        \caption{December 2025 replication of wallet-score persistence and adjusted validation-window markouts. Bars are 95\% confidence intervals.}
        \label{fig:persistence-dec2025}
    \end{minipage}
\end{figure}

\paragraph{The score is persistent and the tail is stable.}
If the July ranking captures a stable wallet attribute, rank persistence should survive in a
different quarter and reconstruction. Across adjacent December windows, Spearman $\rho$ is 0.47,
close to July's 0.52 (Figure~\ref{fig:persistence-dec2025}(a)).

Rank persistence matters because it again translates into an economically distinct upper tail.
The scoring-window decile ladder is monotone, and validation markouts rise sharply at the top
(Figure~\ref{fig:persistence-dec2025}(b)). The top decile's ten-second markout remains stable, and it holds a steady quarter of all
discretionary taker notional across the two windows, 26.7\% and 26.0\%, close to July's
31.0\% and 25.1\% (Table~\ref{tab:part1-panels-dec2025}).

\begin{table}[H]\centering
\begin{threeparttable}
\caption{December 2025 wallet deciles: scoring window and out-of-sample validation}
\label{tab:part1-panels-dec2025}
\footnotesize\setlength{\tabcolsep}{2.9pt}
\begin{tabular}{lrrrrrrrrrr}
\toprule
 & \multicolumn{5}{c}{Scoring window (Dec. 1--10)} & \multicolumn{5}{c}{Validation window (Dec. 11--20)} \\
\cmidrule(lr){2-6} \cmidrule(lr){7-11}
 & Wallets & Orders & Markout & Median & Notional & Wallets & Orders & Markout & Median & Notional \\
Decile & & & (bps) & order (\$) & share (\%) & & & (bps) & order (\$) & share (\%) \\
\midrule
D1 & 252 & 70,235 & $-$1.51 & 136 & 1.0 & 211 & 45,558 & $-$0.07 & 106 & 0.8 \\
D2 & 253 & 101,242 & $-$0.46 & 199 & 1.4 & 208 & 78,485 & 0.21 & 285 & 1.5 \\
D3 & 253 & 139,770 & $-$0.23 & 45 & 1.3 & 206 & 86,758 & $-$0.06 & 84 & 0.8 \\
D4 & 252 & 158,969 & $-$0.08 & 32 & 2.1 & 183 & 103,876 & 0.02 & 13 & 0.4 \\
D5 & 253 & 331,341 & 0.02 & 147 & 1.9 & 192 & 133,651 & 0.10 & 145 & 0.9 \\
D6 & 253 & 184,267 & 0.13 & 84 & 1.5 & 213 & 87,907 & 0.15 & 81 & 0.9 \\
D7 & 252 & 111,494 & 0.30 & 207 & 4.0 & 208 & 70,910 & 0.98 & 131 & 2.7 \\
D8 & 253 & 127,782 & 0.50 & 748 & 5.8 & 221 & 81,911 & 0.21 & 634 & 5.6 \\
D9 & 253 & 187,731 & 1.05 & 989 & 10.4 & 213 & 151,204 & 0.97 & 1,044 & 9.6 \\
D10 & 252 & 1,179,099 & 2.25 & 731 & 26.7 & 240 & 828,640 & 2.34 & 810 & 26.0 \\
\bottomrule
\end{tabular}
\begin{tablenotes}[flushleft]\footnotesize
\item \textit{Notes:} Deciles are formed once on the frozen Dec. 1--10 raw score and never
re-sorted. Both blocks report the same five statistics for qualifying aggressive orders on BTC,
ETH, and SOL (TWAP and liquidation orders excluded): distinct active wallets, order count,
notional-weighted ten-second markout, median order notional, and the decile's share of all
discretionary taker notional in that window. That share column is taken against the whole
window's taker flow rather than the scored subtotal, so it does not sum to one hundred percent;
the residual is flow from addresses without a qualifying scoring-window history.
\item The scoring window is Dec. 1--10 and the validation window Dec. 11--20; the Dec. 21--31
evaluation window is not used here.
\end{tablenotes}
\end{threeparttable}\end{table}

\paragraph{Identity predicts returns, on the same shape.}
If the forecasting gain is not specific to July, identity should again improve the anonymous
benchmark out of sample. Table~\ref{tab:part2-v1-dec2025} confirms this prediction: the identity
block raises
out-of-sample $R^2$ by $+14.6\%$ at the one-second headline horizon ($t=3.9$), with larger gains
at 0.2 and 0.5 seconds. As in July the gain is largest at the shortest horizons; unlike July it
does not decay monotonically, ticking up again at thirty seconds, where the anonymous $R^2$
is below 3\% and a relative gain is not comparable to the short-horizon cells. The anonymous
benchmark is comparable to July's (12.36\% against 10.88\% at one second), so the gain is not an
artifact of a weaker baseline in the replication sample.

\begin{table}[H]\centering
\begin{threeparttable}
\caption{December 2025 return prediction, top decile}
\label{tab:part2-v1-dec2025}
\footnotesize\setlength{\tabcolsep}{4.5pt}
\begin{tabular}{lrrrrrrrr}
\toprule
 & \multicolumn{4}{c}{Ridge (benchmark)} & \multicolumn{4}{c}{Gradient-boosted trees} \\
\cmidrule(lr){2-5} \cmidrule(lr){6-9}
Horizon & Anonymous & + Identity & Gain (\%) & $t$ & Anonymous & + Identity & Gain (\%) & $t$ \\
 & \multicolumn{2}{c}{$R^2$ (\%)} & & & \multicolumn{2}{c}{$R^2$ (\%)} & & \\
\midrule
\multicolumn{9}{l}{\textit{Panel: Frozen score}} \\
\midrule
0.2s & 8.70 & 10.75 & +23.5 & 5.3 & 26.30 & 28.76 & +9.3 & 5.3 \\
0.5s & 12.71 & 15.24 & +19.9 & 4.9 & 24.34 & 28.91 & +18.8 & 5.4 \\
1s & 12.36 & 14.16 & +14.6 & 3.9 & 21.16 & 25.15 & +18.8 & 5.1 \\
2s & 10.50 & 11.58 & +10.3 & 3.8 & 13.84 & 15.10 & +9.0 & 6.3 \\
5s & 8.13 & 8.66 & +6.4 & 3.4 & 9.08 & 9.64 & +6.3 & 7.4 \\
10s & 6.03 & 6.38 & +5.8 & 4.0 & 6.18 & 6.50 & +5.3 & 7.9 \\
30s & 2.66 & 2.86 & +7.8 & 4.8 & 2.68 & 2.80 & +4.4 & 5.0 \\
\midrule
\multicolumn{9}{l}{\textit{Panel: Rolling ten-day score}} \\
\midrule
0.2s & 8.70 & 10.85 & +24.7 & 5.1 & 26.30 & 28.60 & +8.7 & 6.2 \\
0.5s & 12.71 & 15.28 & +20.3 & 4.8 & 24.34 & 28.67 & +17.8 & 6.4 \\
1s & 12.36 & 14.13 & +14.4 & 4.0 & 21.16 & 24.74 & +16.9 & 5.7 \\
2s & 10.50 & 11.50 & +9.6 & 3.8 & 13.84 & 14.52 & +4.9 & 8.4 \\
5s & 8.13 & 8.57 & +5.4 & 2.9 & 9.08 & 9.50 & +4.7 & 8.6 \\
10s & 6.03 & 6.30 & +4.6 & 3.2 & 6.18 & 6.23 & +0.9 & 5.1 \\
30s & 2.66 & 2.79 & +5.0 & 3.4 & 2.68 & 2.76 & +3.2 & 6.2 \\
\bottomrule
\end{tabular}
\begin{tablenotes}[flushleft]\footnotesize
\item \textit{Notes:} The December 2025 counterpart of Table~\ref{tab:part2-v1}, on the
top decile of the score named in each panel. Scores use December 1--10, model fitting December
11--20, and evaluation December 21--31; variable blocks, targets and $R^2$ definitions are
unchanged.
\item \emph{Anonymous} is the benchmark without identity features and \emph{+ Identity}
adds the identity block; both report out-of-sample $R^2$. Gain
$=R^2_{\mathrm{+Identity}}/R^2_{\mathrm{Anonymous}}-1$ is the relative increase, and $t$
uses eleven paired daily MSE differences.
\end{tablenotes}
\end{threeparttable}\end{table}

\paragraph{The matched-cohort comparison holds.}
A genuine toxicity ranking should also remain more informative than wallets matched on size and
activity. Against 200 cohorts drawn on empirical-decile cells, the December toxic cohort exceeds
every draw at every reported horizon and both tail cuts (rank $p=0.005$
throughout; Table~\ref{tab:p2placebo-dec2025}). The best of 200 draws reaches 62 to 83\% of the real
increment across the eight reported cells, a thinner margin than July's 25 to 82\% over the same
eight cells (Table~\ref{tab:p2placebo}), but no draw reaches the real cohort in any cell. Exact-cell
donor limits make the draws smaller than the full toxic cohort, so this rank is descriptive rather
than an equal-size randomization test.

\begin{table}[H]\centering
\begin{threeparttable}
\caption{December 2025 matched-cohort comparison}
\label{tab:p2placebo-dec2025}
\small\setlength{\tabcolsep}{4pt}
\begin{tabular}{lrrrrrrrr}
\toprule
 & & \multicolumn{3}{c}{Actual toxic cohort} & \multicolumn{3}{c}{200 placebo cohorts, $\Delta R^2$ (pp)} & \\
\cmidrule(lr){3-5} \cmidrule(lr){6-8}
Horizon & Anonymous $R^2$ (\%) & $\Delta R^2$ (pp) & Gain (\%) & $t$ & mean & p95 & max & $p$ \\
\midrule
\multicolumn{9}{l}{\textit{Top decile}} \\
\midrule
0.2s & 8.73 & +1.045 & +12.0 & 3.3 & +0.414 & +0.641 & +0.792 & 0.0050 \\
0.5s & 12.78 & +1.876 & +14.7 & 3.8 & +0.779 & +1.192 & +1.448 & 0.0050 \\
1s & 12.38 & +1.720 & +13.9 & 4.0 & +0.731 & +1.122 & +1.364 & 0.0050 \\
10s & 5.92 & +0.226 & +3.8 & 2.3 & +0.073 & +0.144 & +0.180 & 0.0050 \\
\midrule
\multicolumn{9}{l}{\textit{Top quintile}} \\
\midrule
0.2s & 8.73 & +1.072 & +12.3 & 3.9 & +0.409 & +0.615 & +0.660 & 0.0050 \\
0.5s & 12.78 & +1.851 & +14.5 & 4.4 & +0.789 & +1.199 & +1.281 & 0.0050 \\
1s & 12.38 & +1.689 & +13.7 & 4.6 & +0.742 & +1.134 & +1.196 & 0.0050 \\
10s & 5.92 & +0.214 & +3.6 & 2.6 & +0.090 & +0.168 & +0.179 & 0.0050 \\
\bottomrule
\end{tabular}
\begin{tablenotes}[flushleft]\footnotesize
\item \textit{Notes:} Panels give the identity block built from the top decile and the top
quintile of the frozen Dec. 1--10 toxicity ranking, with draws matched on cells crossing
empirical deciles of scoring-window notional and order count. Exact matching is limited by
non-toxic donor supply, so each top-decile (top-quintile) draw contains 154 (388) wallets while
the actual cohort retains all 253 (505); the rank is descriptive rather than an equal-size
randomization test.
\item Ridge models are independently tuned and refit on the 100ms decision grid, with identity
quote state sampled each second and carried forward, so the actual comparator is a
grid-specific refit rather than the headline model of Table~\ref{tab:part2-v1-dec2025}.
\item Gain $=R^2_{\mathrm{+Identity}}/R^2_{\mathrm{Anonymous}}-1$ is the relative
increase, and the placebo columns summarize the same increment
across draws. $p=(1+\#\{\mathrm{placebo}\ge\mathrm{real}\})/201$, so $p=0.0050$ is the floor
attained when no draw reaches the actual cohort. Scores use Dec. 1--10, fitting Dec.
11--20, and evaluation Dec. 21--31; $t$ uses eleven daily MSE differences.
\end{tablenotes}
\end{threeparttable}\end{table}

\begin{figure}[H]
    \centering
    \includegraphics[width=\textwidth]{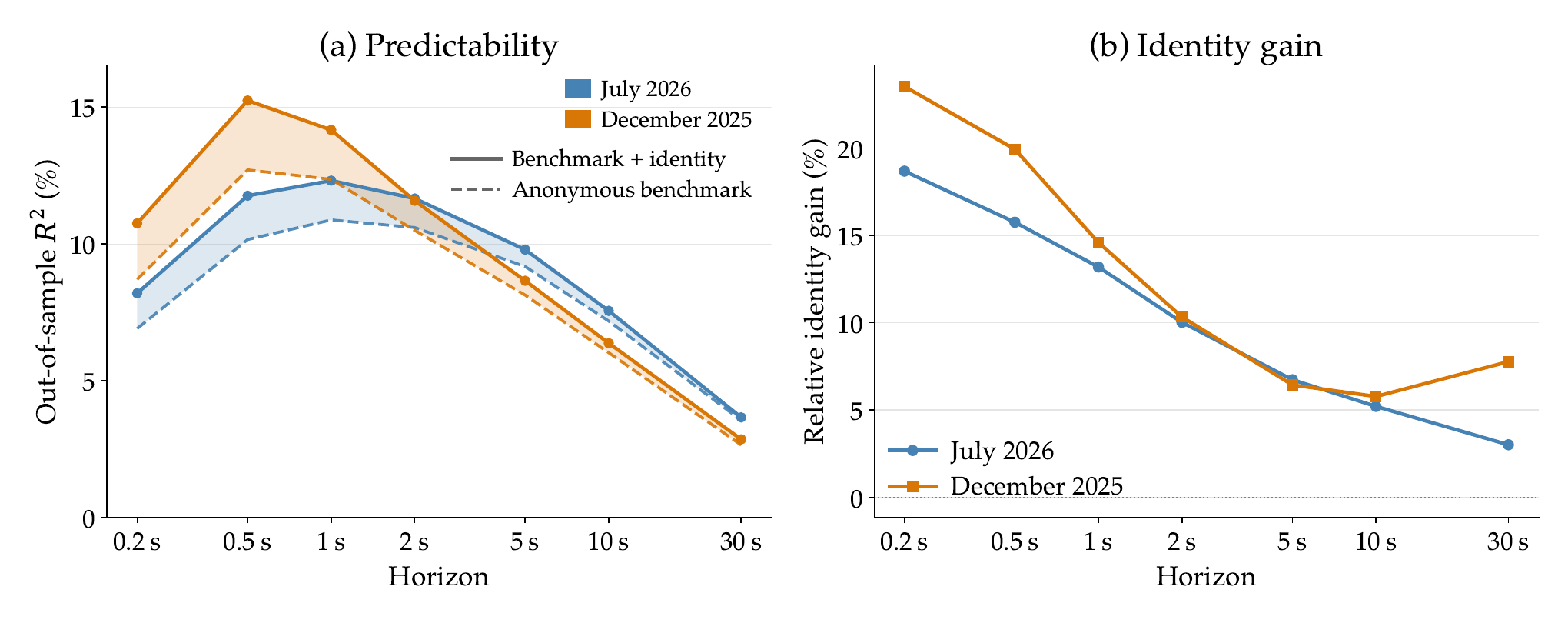}
    \caption{Top-decile ridge results in July 2026 and December 2025. Panel (a) compares the anonymous benchmark with the model including identity; Panel (b) shows the relative identity gain.}
    \label{fig:replication}
\end{figure}

\paragraph{The two samples agree on the shape of the gain.}
December is a different quarter, collected by another party and reconstructed here from the public
release, and the design is applied to it without retuning. The replication reproduces both results: identity raises
out-of-sample $R^2$ at every reported horizon (Table~\ref{tab:part2-v1-dec2025}), and the toxic cohort exceeds every
activity-matched draw (Table~\ref{tab:p2placebo-dec2025}). A common mechanism should also produce
a similar horizon profile. Figure~\ref{fig:replication} confirms this: in both samples, the
ridge identity gain is largest at the shortest horizons and falls monotonically through ten
seconds.
This shared decay is difficult to reconcile with a sample-specific artifact.

\section{How Large Is the Increment?}
\label{app:lean}

A forecasting increment is easier to judge in payoff units than in $R^2$, so we convert the
half-second increment into a midpoint payoff per unit of gross exposure. At each grid step the
position is proportional to the anonymous or identity-augmented forecast. Activation thresholds are
common ex ante budgets based either on trailing-ten-day or fitting-window forecast-magnitude
quantiles; realized acted-grid shares are reported separately.

Per unit of gross exposure, the always-on identity edge exceeds the benchmark's by 7.0\%
($t=7.8$). That is close to the $\sqrt{11.76/10.16}-1=7.6\%$ implied mechanically by the $R^2$
gain, which is the check the exercise has to pass: the conversion should reproduce the forecasting
result in payoff units, and it does. Thresholded comparisons vary across activation budgets
(Table~\ref{tab:lean}).

\begin{table}[tbp]\centering
\begin{threeparttable}
\caption{Frictionless conversion diagnostic by activation budget}
\label{tab:lean}
\footnotesize\setlength{\tabcolsep}{5pt}\renewcommand{\arraystretch}{1.05}
\begin{tabular}{lrrrrr}
\toprule
 & \multicolumn{2}{c}{Payoff (bps)} & & & \\
\cmidrule(lr){2-3}
Budget & Anonymous & + Identity & Increment & Relative gain & $t$ \\
\midrule
\multicolumn{6}{l}{\textit{Panel A: rolling ten-day thresholds}} \\
\midrule
1\% & 0.754 & 0.752 & -0.0012 & -0.2\% & -0.1 \\
5\% & 0.470 & 0.495 & +0.0253 & +5.4\% & \phantom{-}2.9 \\
10\% & 0.351 & 0.371 & +0.0195 & +5.5\% & \phantom{-}4.0 \\
20\% & 0.249 & 0.264 & +0.0144 & +5.8\% & \phantom{-}5.0 \\
\midrule
\multicolumn{6}{l}{\textit{Panel B: thresholds frozen on the fitting window}} \\
\midrule
1\% & 0.757 & 0.755 & -0.0019 & -0.2\% & -0.2 \\
5\% & 0.485 & 0.524 & +0.0388 & +8.0\% & \phantom{-}3.8 \\
10\% & 0.361 & 0.385 & +0.0232 & +6.4\% & \phantom{-}5.3 \\
20\% & 0.255 & 0.271 & +0.0160 & +6.3\% & \phantom{-}6.1 \\
\midrule
\multicolumn{6}{l}{\textit{Panel C: no threshold (always active)}} \\
\midrule
Always active & 0.131 & 0.140 & +0.0091 & +7.0\% & \phantom{-}7.8 \\
\bottomrule
\end{tabular}
\begin{tablenotes}[flushleft]\footnotesize
\item \textit{Notes:} Payoff is the realized half-second midpoint payoff per unit of gross
exposure, $\sum \hat y\,y/\sum|\hat y|$ over acted moments (bps), July 21--27. Increment is
the \emph{+ Identity} minus \emph{Anonymous} payoff difference in the same units, and
Relative gain expresses it as a share of the \emph{Anonymous} payoff.
\item Thresholds are per-coin $(1-b)$-quantiles of $|\hat y|$ over the trailing ten days
(Panel~A) or the July 11--20 fitting window (Panel~B); out-of-sample gates need not fire at the
nominal budget, and the realized share of grid steps acted on is at or below it in every cell.
$t$ comes from the seven paired daily differences, consecutive decisions overlapping within the
horizon.
\end{tablenotes}
\end{threeparttable}\end{table}

\noindent The exercise is frictionless and evaluated at the midpoint, with no book, queue, fill
probability, spread, or inventory, so it is a unit conversion rather than evidence of capture. As a
standalone taker strategy the gated rows' 0.25--0.76 basis points per unit of gross exposure would
not cover the venue's 4.5-basis-point median taker fee. That comparison is the wrong benchmark for
the mechanism the paper documents: the participant who can use the signal is a maker already
quoting, for whom the relevant quantity is adverse selection avoided rather than a fee paid.

\end{document}